# Quantifying the Dynamics of Network Detachment Across Scientific, Technological, Commercial, and Pharmacological Domains

Binglu Wang[1,2,3*], Ching Jin[1,2,3,4*], Chaoming Song[5], Johannes Bjelland[6], Brian Uzzi[1,2,3] & Dashun Wang[1,2,3†]

[1] *Center for Science of Science and Innovation, 1800 Sherman Avenue, Evanston, IL 60201, USA*

[2] *Northwestern Institute on Complex Systems (NICO), 600 Foster Street, Evanston, IL 60208, USA*

[3] *Kellogg School of Management, 2211 Campus Drive, Evanston, IL 60208, USA*

[4] *Centre for Interdisciplinary Methodologies, University of Warwick, Coventry, CV4 7AL, United Kingdom*

[5] *Department of Physics, University of Miami, 1320 S Dixie Hwy, Coral Gables, FL 33146, USA*

[6] *Telenor Research and Development, Snarøyveien 30 N-1360 Fornebu, Norway*

[*] *These authors contributed equally to the work.*

[†] *Correspondence should be addressed to D.W. (dashun.wang@kellogg.northwestern.edu)*

**Despite the vast literature on complex network dynamics, our understanding of the network detachment process remains limited. This understanding is essential for deeper insight into network evolution. Here, we analyze four large-scale datasets that capture the structural patterns of detachment across scientific, technological, commercial, and pharmacological domains. Across these diverse domains, we uncover one simple pattern of preferential detachment in social systems, whereby the probability for individuals or organizations to disengage from a previously adopted practice correlates with the number of network neighbors who have already detached. To examine the structural consequences of this empirical pattern, we show that preferential detachment can fundamentally alter how networked systems disintegrate, inducing a phase transition in systems typically considered robust against detachment. We further derive an analytical framework to systematically understand the impact of preferential detachment on network dynamics, pinpointing the topological conditions under which it accelerates, decelerates, or leaves unchanged the breakdown of the network compared with random detachment. Together, these results demonstrate that the detachment dynamics follow simple yet reproducible patterns, deepening our quantitative understanding of detachment within networked social systems, with implications for the robustness and functioning of innovation communities.**

Over the past several decades, research on complex networks has transformed our understanding of how interconnected systems form and expand across a diverse range of domains, from scientific collaboration and technological adoption to online information diffusion [1-14]. A fundamental property of many networked systems is that popularity or activity eventually decays over time [15-19]. While the growth of networks has been extensively studied, the counterpart of this process—how networks disconnect through the disengagement or detachment of nodes—has received relatively little attention, especially within social contexts [16, 20], leaving open important questions about the dynamics of network detachment.

Existing research on network breakdown largely stems from studies of non-social systems such as power grids, the Internet, financial infrastructures, and ecological food webs [2, 3, 21, 22]. In these domains, node removal is typically modeled as resulting from random failures, targeted attacks, or optimal percolation strategies, with a focus on structural robustness and systemic resilience [23, 24]. While these frameworks provide valuable insights, they abstract away from a defining feature of social systems: detachment is rarely random. Instead, real-world social systems may exhibit critical network dynamics that arise from socially embedded decisions, shaped by peer behavior, structural interdependencies, and shifting incentives [25]. Such socially mediated detachment can propagate through networks, producing fragmentation or even systemic collapse, yet its structural regularities remain largely understudied.

One key reason for this gap in social systems is the historical lack of large-scale datasets that systematically capture the detachment dynamics of networks [16]. Disengagement decisions are rarely recorded with the same precision as adoption or engagement events, making it difficult to systematically analyze patterns of detachment. To address this limitation, we compile and analyze four large-scale datasets spanning the scientific, technological, commercial, and pharmacological domains.

The first dataset ($D_1$) captures the detachment dynamics of scientific fields as scientists shift their focus away from certain areas of research. Here we use the Microsoft Academic Graph data [26] to measure the fraction of publishing scientists in each field over time, allowing us to delineate the life cycle of 1,219 scientific fields from 1940 to 2015 and quantify the dynamics of scientists leaving a field (SI S1.1). Our second dataset ($D_2$) records the dynamics of inventors shifting across patenting domains, including 6.9M patents granted by the USPTO from 1976 to 2015. Similar to $D_1$, here we trace the life cycles of 291 patenting domains, approximated by Cooperative Patent Classification (CPC) codes, and measure the proportion of inventors in each domain over time (SI S1.2). For $D_1$ and $D_2$, we also construct the collaboration network using co-authorship and co-inventor information, respectively [4].

Our third dataset ($D_3$) focuses on a commercial domain and traces the individual usage history of mobile handsets within a European country (704 types of handsets by 3.5M users), allowing us to quantify the detachment dynamics of mobile handsets through changes in daily active users over a period of nine years (2006-2014). In addition, we collect anonymized phone call records among 1.7M users within a three-month period (October 1, 2013-January 1, 2014) to approximate the social network among a subset of users for the network-based analyses (SI S1.3). Lastly, we construct our fourth dataset ($D_4$) tracing the dynamics of pharmacological technologies and analyze the detachment of drug mechanisms, i.e., biological mechanisms of action (MOAs), as pharmaceutical organizations shift their focus away from certain MOAs in drug development. Here we use the Cortellis Drug Discovery Intelligence database, which contains the largest and most comprehensive drug development records in the world, covering 648,702 drugs over the past 25 years [27, 28]. To quantify the detachment dynamics of pharmacological technologies, we examine MOAs associated with each drug and measure the R&D focus of pharmaceutical companies on these MOAs. In total, we trace 248 MOAs by 5,133 pharmaceutical companies and measure the fraction of companies using each MOA in drug development. We also calculate the similarity in market focus among pharmaceutical companies to approximate the underlying proximity between two companies (SI S1.4). The item-specific relational networks exhibit heterogeneous, fat-tailed degree distributions across all four empirical domains (SI S2.1). Together, these four datasets offer us an unprecedented opportunity to examine the temporal life-cycle dynamics and the structural pattern in which the detachment processes unfold.

Figure 1A,D,G,J plots the popularity dynamics of different scientific fields, patenting domains, mobile handsets, and drug MOAs, respectively. All four domains exhibit a similar rise-and-fall pattern, demonstrating that both the adoption and detachment processes are important for understanding these life-cycle dynamics. Here we focus on the declining phase of each item and measure the rate of detachment $\nu$ as a function of time $(t)$. We find that, across scientific fields,

patenting domains, mobile handsets, and drug MOAs, the average detachment probability ($\nu$) systematically increases with time ($t$) (Fig. 1B,E,H,K). Going beyond the aggregate estimates, we test if the observed trend holds for individual scientific fields, patenting domains, mobile handsets, or drug MOAs. In particular, for each item, we build panel data containing all of its users, fit this panel data using linear regression, and estimate the slope of the fit for each item, to approximate how its detachment rate changes as a function of time (Fig. 1C,F,I,L). We find that the majority of scientific fields (98%), patenting domains (92%), mobile handsets (91%), and drug MOAs (84%) can be approximated by a positive slope, suggesting the detachment probability of different items appears to trend upward in a rather ubiquitous manner.

To elucidate the relationship between detachment rates and underlying network structures, we further test the hypothesis that detachment processes follow Poissonian dynamics. This is a prevailing assumption in computational models, often derived from epidemiological frameworks [5, 29-37], and it has shown initial success in modeling substitutive systems [5]. This assumption reflects the idea that items follow a characteristic lifespan and that the probability of detachment is independent of the behavior of an individual's network neighbors. To test this implication, we control for potential temporal effects by restricting our measurements to a short time interval beginning around each item's half-life (i.e., the stage at which its popularity has fallen to half of its peak), where detachment turnover is most active and a common life-cycle stage is shared across items with different lifespans (Fig. 2; SI S2.2). We verify that our results are robust to this choice of time window (SI S2.3). For each item $i$, we consider its item-specific relational network, namely the subnetwork of the underlying relational network formed by users associated with $i$, and divide users into two groups depending on whether detachment events have occurred among their network neighbors ($r_i > 0$ vs. $r_i = 0$) within the observation window. We find that the distributions of detachment rates for the two groups show visibly different patterns (Fig. 2A-D), with the distribution for the first group ($r_i > 0$) shifting systematically to the right of the $r_i = 0$ group. These findings suggest that the detachment rate tends to be higher when detachment events have occurred among an individual's network neighbors. We then test whether the detachment probability ($\nu_i$) increases with the number of network neighbors who have also detached from the item ($r_i$). Figure 2E-H shows four case studies, where we measure the rate of detachment as a function of $r_i$ for one selected example in each dataset. Averaged curves across all items in each domain show similar upward trends, confirming that the representative examples are consistent with population-level patterns (SI S2.4). Across these four cases, $\nu_i$ shows a positive relationship with $r_i$, indicating that the detachment probability appears to increase with the number of network neighbors who have detached. By contrast, in a null model of random detachment, $\nu_i$ is independent of $r_i$ (Fig. 2E-H, grey dashed line). To systematically test across different items in different domains (i.e., scientific fields, patenting domains, mobile handsets, and drug MOAs), we measure $\nu_i$ and $r_i$ for each item in the four datasets, and we fit them with a linear ansatz, finding that a substantial fraction of scientific, technological, commercial, and pharmacological items exhibit a positive slope (73%, 70%, 79%, and 70%, respectively) (Fig. 2I-L). This pattern also remains robust when exposure is normalized by degree, $r_i/k_i$ , indicating that the result is not driven solely by the raw number of detached neighbors (SI S2.5).

Overall, these empirical findings reveal a systematic pattern that we refer to as preferential detachment: the probability that an individual disengages increases with the number of detachments among their network neighbors. This observation contrasts with the Poissonian assumption that underlies most prevailing computational frameworks for modeling network detachment dynamics. We use the term "preferential" in a statistical sense, rather than to imply conscious preference or a specific causal mechanism. By analogy to preferential attachment, it denotes a non-uniform detachment probability with respect to a structural quantity, here local exposure to prior detachment among network neighbors. We also distinguish this usage from prior uses of "preferential detachment" in network growth and rewiring models, where the term often refers to degree-dependent edge removal, link rewiring, node deletion, or the loss of structural connections [38-41]. Here detachment is defined as item-specific disengagement conditioned on neighbors' detachment states. Most importantly, the scientific, technological, commercial, and pharmacological domains we studied differ substantially in their scope, scale, temporal resolution, and nature. Yet despite these clear differences, and the myriad factors that might affect the decisions of individuals and organizations to detach from an innovative pursuit [42-47], the patterns of preferential detachment show remarkable consistencies across these rather diverse domains.

The concept of preferential detachment challenges the Poissonian assumption that has traditionally underpinned computational models of network decline. At the same time, it connects to several important ideas in social science, from peer effects and network externalities [44, 48-52] to the theory of creative destruction [53] to vacancy chains [47, 54]. Interestingly, preferential detachment mirrors preferential attachment [1, 55-58], the well-established mechanism governing network growth that generates scale-free structures with heterogeneous degree distributions. If preferential attachment governs how networks grow, preferential detachment may help explain how they decline. This parallel raises a natural question: what structural consequences might follow from this empirically observed detachment pattern?

To examine one such consequence, we study how preferential detachment reshapes network structure during periods of decline. We incorporate this empirical pattern into a minimal network model. Specifically, we approximate disengagement as node removal and formalize a detachment process on a network with $N$ nodes. At each time step, node $i$ is removed with probability:

$$\pi_i = \frac{\alpha r_i + (1-\alpha)}{\sum_{j \in S} \alpha r_j + (1-\alpha)}, \tag{1}$$

where $r_i$ denotes the number of neighbors of node $i$ that have already been detached, $\alpha$ captures the strength of preferential detachment, and $S$ denotes the set of surviving nodes. When $\alpha = 0$, nodes are removed randomly with the same probability, and Eq. (1) recovers the existing modeling framework of random detachment. As $\alpha$ increases, node removal becomes increasingly conditioned on local detachment in the network. Estimating an empirical analogue of the preferential-detachment strength further shows positive item-level $\alpha$ values across all four domains (SI S2.6), providing a direct empirical bridge between the observed exposure–detachment relationship and the model parameter.

To analytically investigate the process, we adopt the heterogeneous mean-field approximation (HMA) and derive a self-consistent condition for the critical threshold ($f_c$) under preferential detachment. The critical point satisfies:

$$f_c = 1 - \frac{<k> \sum_k P_k h(f_c)^k}{\sum_k P_k (k^2 - k) h(f_c)^{k-2}} = 1 - \frac{G_0(h(f_c)) <k>}{G_0''(h(f_c))}, \tag{2}$$

where $G_0(x) \equiv \sum_k P_k x^k$ and $G_1(x) \equiv \sum_k \frac{1}{<k>}(k+1)P_{k+1}x^k$ are the standard generating functions of the degree distribution. The auxiliary function $h(f)$, which captures the evolving local survival probability of edges during detachment, follows the differential equation:

$$\frac{dh}{df} = \frac{-\alpha h\, G_0^2(h) + \alpha\, G_0(h)G_1(h)(1-f)}{\alpha\,(1-f)h\, G_0(h)G_0'(h) - \alpha\,(1-f)^2 G_0'(h)G_1(h) + (1-\alpha)(1-f)G_0^2(h)}, \tag{3}$$

with initial condition $h(f_0)$=1.

Together, Eq. (2) and Eq. (3) provide a self-consistent percolation threshold under preferential detachment. When $\alpha = 0$, preferential detachment reduces to random detachment and Eq. (2) collapses to the canonical Molloy–Reed criterion. For scale-free networks with $\gamma < 3$, the canonical criterion predicts no nontrivial phase-transition point below complete removal when expressed as the removed-node fraction, $f_c \to 1$. The situation changes qualitatively under preferential detachment ($\alpha > 0$). As detachment progresses, $h(f)$ drops below 1, which makes the factor $\sum_k P_k(k^2 - k)h(f_c)^{k-2}$ converge and thereby yields a finite critical $f_c < 1$ satisfying Eq. (2). Thus, regardless of how weak the preference is, preferential detachment restores a genuine phase transition in scale-free networks. In other words, even an infinitesimal amount of preferential detachment fundamentally changes the breakdown dynamics: a phase transition emerges that is absent in the random-detachment case.

To verify this theoretical prediction, we first simulate the detachment process on a specific scale-free network ( $P_k \sim k^{-\gamma}$, with $\gamma = 2.5$), and we study its robustness to random detachment ($\alpha = 0$) and preferential detachment ($\alpha = 0.5$). Specifically, we trace the size of the giant component ($S$) as a function of the fraction of nodes removed ($f$). We find that under random detachment (Fig. 3A, grey line), the network shows a high degree of robustness [2, 29, 37], as its giant component remains connected until all nodes are removed. If, however, the system undergoes preferential detachment (Fig. 3A, red line), a fundamentally different pattern emerges. First, conditional on removing the same number of nodes, the network breaks down much more quickly than under random detachment. More importantly, the network appears to undergo a critical process and become fragmented when the removed node fraction exceeds a percolation threshold ($f_c$).

We further simulate the preferential detachment process with different $\alpha$ (Fig. 3B), and simulation results show excellent agreement with our analytical calculations derived from Eq. (2). To further illuminate the process, we visualize the network when half of the nodes are removed ($f = 0.5$) for both processes (subpanels in Fig. 3A). While the two networks have the same number of nodes remaining, they show markedly different patterns. Whereas a giant component composed of high-degree nodes persists under random detachment ($\alpha = 0$), the giant component in the preferential detachment case ($\alpha = 0.5$) only contains a small number of low-degree nodes. This highlights a key difference between random and preferential detachment on scale-free networks: compared with random detachment, hubs become disproportionately vulnerable under preferential detachment.

These results align closely with the network science literature [2, 5, 16, 29, 37, 55-69], which has extensively examined the importance of hubs in maintaining the robustness of networks. When hubs are deliberately targeted, the network collapses early (Fig. 3B, dotted line). This raises a natural question: is preferential detachment equivalent to a targeted hub-attack mechanism? To answer this question, we further examine preferential detachment on two additional network topologies: Erdős–Rényi (ER) networks with a Poisson degree distribution and a Gaussian network with a normal degree distribution. If preferential detachment is merely a hub attack, a similar earlier onset of phase transition should be expected for both networks (Fig. 3E,H, black dotted line) [2, 56, 65].

However, Eq. (2) predicts different behavior. Specifically, for ER networks, $G_0(x) = G_1(x) = e^{<k>(x-1)}$. Substituting this into Eq. (2) yields the canonical Molloy–Reed criterion. In other words, preferential detachment follows indistinguishable dynamics from random detachment for the ER network. Interestingly, for Gaussian networks with degree distributions narrower than Poisson, preferential detachment induces an even later percolation transition than random detachment (Fig. 3G), where the critical point is delayed as $\alpha$ increases (Fig. 3H). Both predictions are further confirmed by simulation results (Fig. 3D-I). These results indicate that preferential detachment is not a targeted hub-attack mechanism. Instead, it has highly non-trivial effects on different network topologies, highlighting a novel form of critical behavior and motivating a closer examination of the underlying mechanism.

To further understand how the percolation threshold shifts for different network topologies, we examine the relationship between a node's number of detached neighbors ($r$) and its current degree ($k_f$) after a fraction $f$ of nodes has detached. We first measure $k_f$ vs. $r$ at different levels of $f$ for the scale-free network (Fig. 3C), finding that the two quantities follow an asymptotically linear relationship, suggesting that nodes with larger $r$ are also characterized by a higher remaining degree, mimicking the attack processes (see SI S4 for details). However, when we repeat the analyses on ER networks, we find that $k_f(r)$ remains constant for any given $r$ (Fig. 3F) under preferential detachment, as it does under random detachment. Because connections in ER networks are formed randomly, the number of a node's detached neighbors is statistically independent of its remaining degree during the detachment process. While nodes with higher $r$ are more likely to be selected under preferential detachment, the corresponding $k_f$ is random for ER networks, explaining why preferential detachment on ER networks follows the same dynamics as random detachment.

Mathematically, the asymptotic behavior of $k_f(r)$ as $r \to \infty$ can be directly calculated by our model:

$$k_f(r) \sim (r+1)\frac{P_k(r+1)}{P_k(r)}, \tag{4}$$

suggesting that this key relationship is governed by the original degree distribution of the network $P(k)$ (see SI S4 for details). For the standard degree distribution considered here, whose tails are eventually decreasing in the large-$r$ regime, we have $\frac{P_k(r+1)}{P_k(r)} < 1$. Eq. (4) therefore predicts that $k_f(r)$ is asymptotically bounded by a linear scaling. For example, it saturates to the linear scaling bound for scale-free networks. However, for the Poisson degree $P(k) \sim \frac{\lambda^k}{k!}$, Eq. (4) predicts $k_f \sim \lambda$, indicating $k_f$ does not change with $r$. Thus, the Poisson degree distribution serves as a key division here. If the degree distribution has a heavier tail than the Poisson distribution, such as power-law distributions, $k_f$ increases with $r$. Hence, in such networks, hubs are more likely to be removed under preferential detachment, leading to an earlier onset of the phase transition. By contrast, for networks with a narrower degree distribution than the Poisson distribution, such as normal distributions, there is a negative correlation between $k_f$ and $r$ (Fig. 3I), where once nodes with higher $r$ are removed, preferential detachment actually targets nodes with lower current degrees, thus inducing an even slower percolation than random detachment. Table 1 summarizes the predictions of Eq. (4) for several common networks.

Table 1 Summary of different network structures

| Network | Degree Distribution | $k_f(r)\|_{r\to\infty}$ |
|---|---|---|
| scale-free networks | $k^{-\gamma}$ | $r$ |
| ER networks | $\frac{\lambda^k e^{-\lambda}}{k!}$ | $\lambda$ |
| Gaussian networks | $e^{-\frac{1}{2}\left(\frac{k-\mu}{\sigma}\right)^2}$ | 0 |

Taken together, this paper examines the declining dynamics across four distinct domains and makes three primary contributions. First, it empirically documents a simple yet ubiquitous phenomenon of preferential detachment governing the dynamics of network detachment. Across scientific, technological, commercial, and pharmacological domains, the probability of detachment is consistently associated with local exposure to prior detachment among network neighbors. These findings provide the central empirical contribution of the paper and suggest that detachment in social and innovation systems is not merely an individual-level or item-level life-cycle process; it is also shaped by local network structure. Second, building on this empirical regularity, we incorporate preferential detachment into current modeling frameworks and find that its presence creates novel yet complex dynamics in the breakdown of the overall ecosystem, generating a new form of structural collapse in networked systems that are thought to be robust against random detachments. Third, we derive an analytical framework to systematically understand the impact of preferential detachment on network breakdown, pinpointing specific conditions under which it accelerates, decelerates, or leaves unchanged the breakdown of the network compared with random detachment, depending on the network topology. Overall, these results provide a new quantitative basis for understanding network detachment in social systems, which has implications for the robustness and functioning of interconnected communities.

While this paper synthesizes empirical evidence from four distinct domains to document the phenomenon of preferential detachment and its impact on the disintegration of networked communities, it is important to acknowledge a significant limitation: the study does not establish the origins of preferential detachment and its potential interplay with preferential attachment. For example, it is possible that the observed preferential detachment is an extension or byproduct of the preferential attachment mechanisms that govern the growth phase of the new communities. Although the findings of this paper remain valid irrespective of the specific origins of preferential detachment, further research into the connections between the growth and decline of networked communities could pave the way for an integrated framework for the whole life cycle of network dynamics, which would deepen our understanding of and ability to predict how networked communities evolve and dissolve.

Further, analyzing data across four diverse domains, this paper mainly focuses on universal detachment patterns that appear consistent across domains. At the same time, there are substantial heterogeneities both within and across domains. For example, scientists often have considerable autonomy in selecting research fields, whereas inventors typically operate under firm-driven managerial strategies. While both scientists and inventors act as originators of ideas, mobile handset users function primarily as consumers. In contrast, the drug-development dataset reflects organizational (firm-level) decisions, where companies shift their R&D focus across different mechanisms of action.

The four domains also differ in the nature of their underlying relational networks. Scientific fields and patenting domains are based on collaboration networks, and mobile handsets are based on direct communication ties, whereas drug MOAs are based on similarity in market focus among pharmaceutical organizations. The persistence of preferential detachment across these different relational settings suggests a common exposure-dependent pattern, while variation in the estimated strength of detachment across domains indicates that the degree of "stickiness" may depend on the type of actors, ties, and institutional context.

Observing consistent detachment patterns across these contextually distinct domains highlights important underlying commonalities in how detachment unfolds and further suggests fruitful directions for future research. For example, which factors tend to be associated with a higher or lower $\alpha$ parameter? Beyond the universal patterns presented in this paper, insights into the heterogeneity in the detachment processes may further our understanding of the potential mechanisms behind the detachment dynamics.

Our framework also connects to recent work on dynamical processes in complex networks, including studies of strategy evolution on dynamic networks and the evolution of prosocial behavior in multilayer populations [70-72]. These studies show how network architecture, temporal variation, and multilayer coupling can shape microscopic behavioral dynamics. Our work is complementary to this literature. Rather than examining how strategies, behaviors, or ties co-evolve on a fixed, multilayer, or time-varying network, we focus on item-specific detachment: how actors disengage from a focal scientific field, patenting domain, handset model, or drug mechanism of action as local detachment accumulates around them. This distinction is also important when relating our framework to temporal rich-club structure. In much of the temporal-network literature, an edge represents a social, communication, or interaction tie, and a hub refers to an actor who is highly connected in the broader temporal network. In our setting, by contrast, the empirical analyses are conducted on item-specific relational networks: subnetworks of the underlying relational network formed by actors associated with a focal item. Therefore, a hub in our analysis is not necessarily a socially central actor in the general network, but an actor with many neighbors within the item-specific participation network. Likewise, detachment does not imply disappearance from the broader network; it means leaving a focal field, domain, product, or mechanism of action. Temporal rich-club stability and preferential detachment therefore capture different levels of network dynamics. Future work could further examine this relationship by stratifying detachment dynamics by degree class, k-core level, and rich-club membership, and by extending the present framework to explicitly multilayer or time-varying network settings.

Overall, this paper demonstrates that the dynamics of detachment across scientific, technological, commercial, and pharmacological domains follow simple yet reproducible patterns, commonly governed by preferential detachment. The phenomenon of preferential detachment fundamentally alters the breakdown of networked ecosystems, with implications for the robustness and functioning of the underlying systems. Curiously, preferential detachment resembles behavior observed across a wide range of settings, including organizational strategy [43, 46], culture [45], financial markets [44], and healthcare [47], suggesting that the concept of preferential detachment may have broad relevance that goes beyond these examples.

**Data and Code Availability**

The scientific publication data used in this work are from Microsoft Academic Graph. The raw data were publicly available at https://www.microsoft.com/en-us/research/project/microsoft-academic-graph. The patent data used in this work are from PatentsView, which is publicly available at https://patentsview.org/download/data-download-tables. The mobile phone data are not publicly available because they contain commercially sensitive information, but they are available from the corresponding author upon reasonable request. This work also uses drug data sourced from Cortellis, and researchers who wish to access these raw data should contact Cortellis directly.  Data and code necessary for reproducing the main results are available at https://github.com/BW-KH/preferential-detachment.

**Acknowledgments**

We thank all members of the Kellogg Center for Science of Science & Innovation (CSSI) and the Northwestern Institute on Complex Systems (NICO) for helpful discussions. D.W. is supported by the Air Force Office of Scientific Research under award numbers FA9550-17-1-0089 and FA9550-19-1-0354, National Science Foundation grant SBE 1829344, the Alfred P. Sloan Foundation G-2019-12485, and Peter G. Peterson Foundation 21048. C.J. acknowledges support from UK Research and Innovation/ESRC through the Metascience Research Grants opportunity OPP569 [grant number UKRI1090]. The funders had no role in study design, data collection and analysis, the decision to publish, or the preparation of the manuscript.

**Author Contributions**

B.W., C.J., C.S., B.U., and D.W. designed the research; B.W. and C.J. collected data and conducted empirical and analytical analyses with help from D.W., C.S., J.B., and B.U.; all authors discussed and interpreted results; B.W., C.J., and D.W. wrote the manuscript; all authors edited the manuscript.

**Competing Interest Statement**

The authors declare no competing interests.

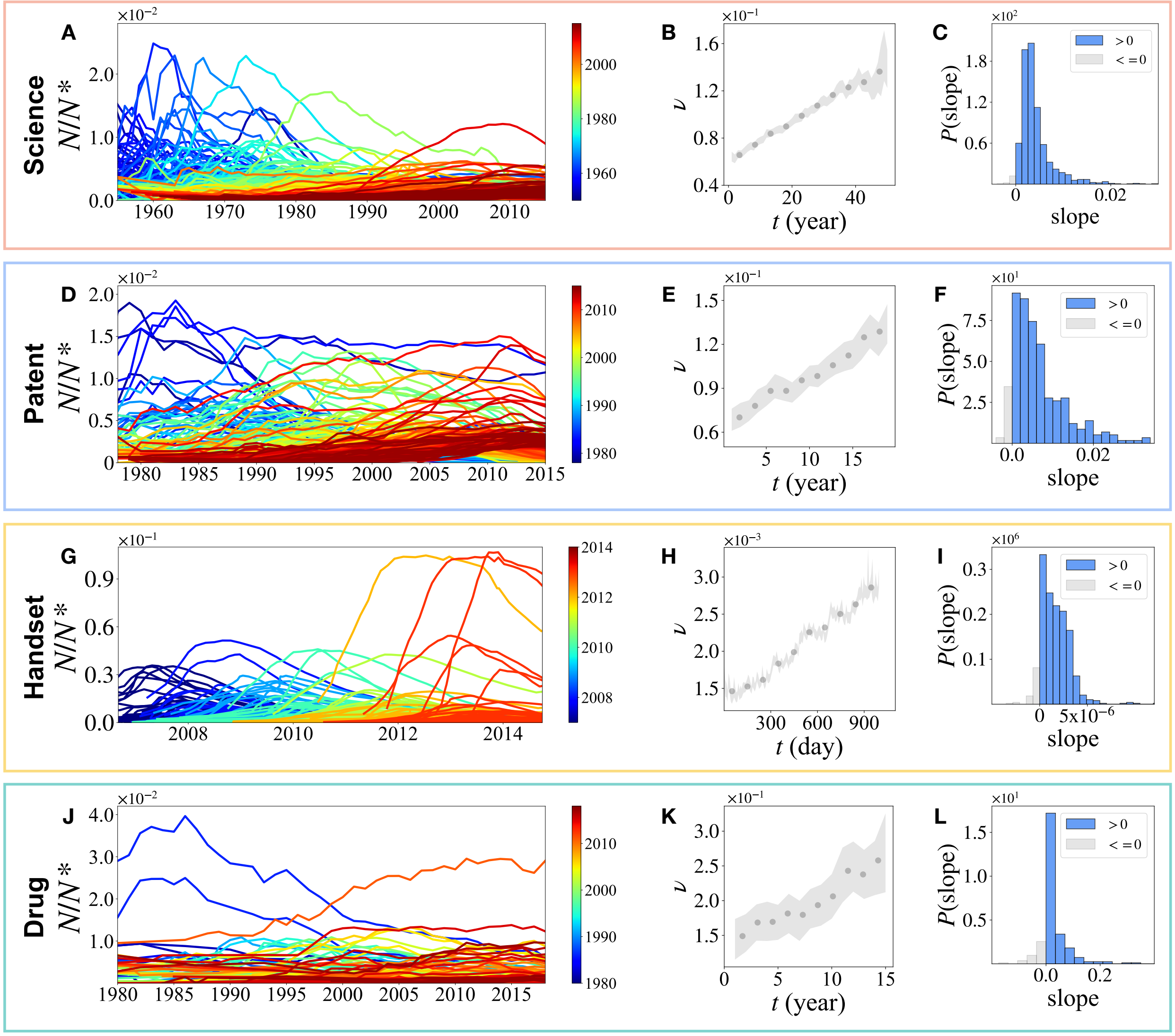

Figure 1: **Temporal dynamics of innovation detachment. (A, D, G, J)** Popularity dynamics ($\frac{N}{N^*}$) of scientific fields (A), patenting domains (D), mobile handsets (G), and drug MOAs (J). Here we calculate the proportion of scientists and inventors in each field, the ratio of active users for each mobile handset, and the fraction of pharmaceutical organizations focusing on each MOA. Each line represents a scientific field (A), patenting domain (D), handset type (G), or drug MOA (J). The color code corresponds to peak dates (when the population reaches its maximum), shifting from blue to red. **(B, E, H, K)** The average detachment probability $\nu \equiv \frac{\Delta N^-}{N}$ as a function of time $t$ for all scientific fields (B), patenting domains (E), mobile handsets (H), and drug MOAs (K), showing that the detachment rate, on average, increases with time. The shaded region represents a 95% confidence interval. **(C, F, I, L)** We estimate the relationship between the detachment probability and time for individual scientific fields (C), patenting domains (F), mobile handsets (I), and drug MOAs (L), and we plot the distributions of the estimated slope coefficients.

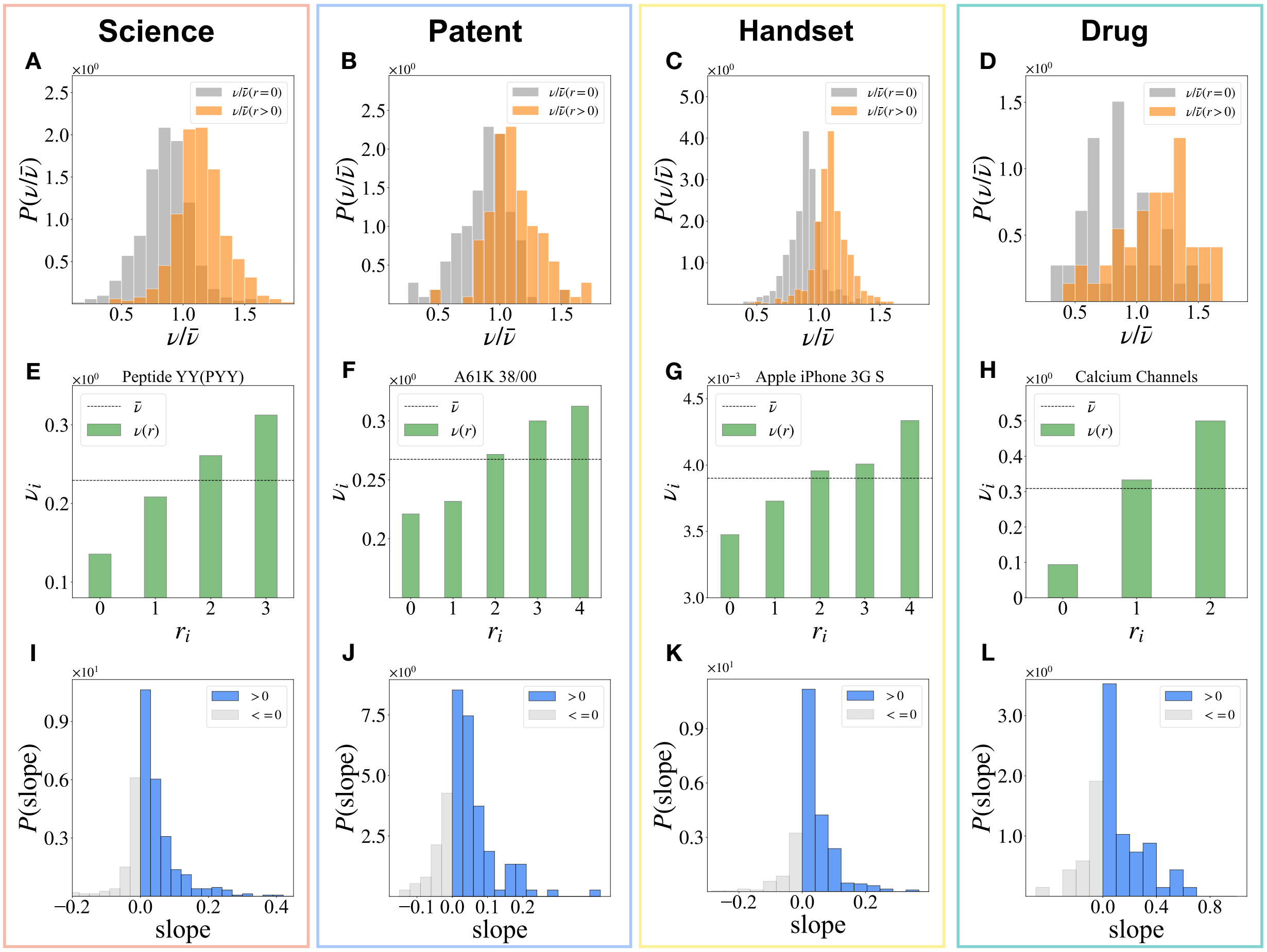

Figure 2: **Detachment and the underlying social network. (A-D)** Distributions of detachment probability plotted separately for people whose network neighbors have detached (orange) or not (grey). Here we control for temporal effects by taking a short time interval for our measurements (one year for the scientific fields, patenting domains, and drug MOAs, and 100 days for the mobile handsets). To maintain consistency of measurements across the four domains, we begin our measurements around the half-life of each life cycle, i.e., when the item's popularity falls to half of its peak, which provides a comparable life-cycle stage across items with different absolute lifespans; robustness checks using alternative reference points are reported in SI S2.3. We find the two distributions deviate from each other, suggesting that the underlying network further plays a role in detachment dynamics. **(E-H)** Detachment probability $\nu_i$ as a function of $r_i$ for one case selected from each domain as an illustrative example (*Peptide YY* for scientific fields, *A61K 38/00* for patenting domains, *iPhone 3GS* for handsets, and *Drugs Targeting Calcium Channels (Voltage-Gated)* for drug MOAs). The four cases were selected as recognizable illustrative examples with sufficient observations across multiple values of $r_i$ for stable visualization. The dashed line indicates the null model, where detachment probability is independent of how many

network neighbors have detached ($r_i$). **(I-L)** We estimate the relationship between the detachment probability $\nu_i$ and $r_i$ for all scientific fields (I), patenting domains (J), mobile handsets (K), and drug MOAs (L), and we plot the distribution of the coefficients, finding that a vast majority of them can be approximated by a positive slope.

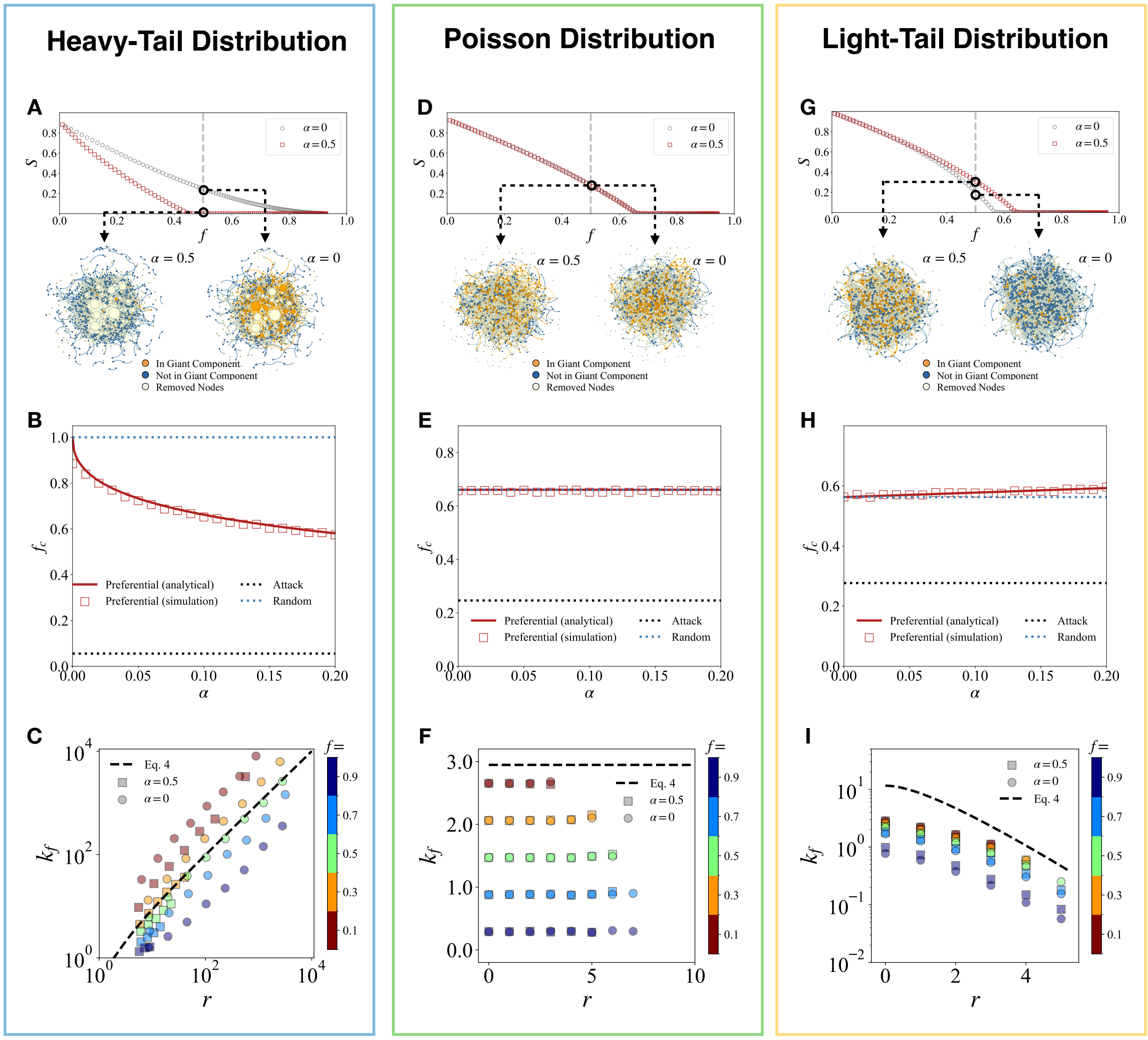

Figure 3: **Preferential detachment across different network topologies.** Simulations of detachment dynamics on networks with different degree distributions: a scale-free network with a heavy-tailed degree distribution (exponent $\gamma = 2.5$; A-C), a network with a Poisson degree distribution (D-F), and a light-tailed network with a normal degree distribution (G-I), all with the same average degree. Simulations are performed on networks of size $N = 10^6$ and averaged over 20 realizations.

**(A, D, G)** Size of the giant component $S$, measured as the fraction of remaining nodes, as a function of the fraction of nodes removed ($f$), under preferential detachment ($\alpha = 0.5$), and under random detachment ($\alpha = 0$). Network visualizations show each network when half of its nodes have detached ($f = 0.5$) and under preferential ($\alpha = 0.5$) and random ($\alpha = 0$) detachment. Here the nodes that have detached are shown in grey; among the remaining nodes, those in the giant connected component are shown in orange, and all others are shown in blue.

**(B, E, H)** Critical fraction of node removal as a function of $\alpha$ for networks with heavy-tailed (B), Poisson (E), and light-tailed (H) degree distributions. For scale-free networks ($\gamma = 2.5$), simulated results under preferential detachment are shown as red squares, with the corresponding analytical prediction shown as a solid orange line. Results for targeted attacks on scale-free networks (black dotted line) and random detachment (blue dotted line) are shown for comparison. We repeat the same analyses for networks with a Poisson distribution in (E) and a light-tailed degree distribution in (H).

**(C, F, I)** Relationship between the current degree $k_f$ and the number of detached neighbors $r$ at different time snapshots ($f = 0.1, 0.3, 0.5, 0.7, 0.9$), obtained from Monte Carlo simulation under random detachment (circles) and preferential detachment ($\alpha = 0.5$; squares) on heavy-tailed (C), Poisson (F), and light-tailed (I) networks. The dashed black line indicates the asymptotic behavior predicted by Eq. (4) in the large-$r$ regime (see SI S4 for details).

## References

1. Barabási, A.-L. and R. Albert, *Emergence of scaling in random networks.* Science, 1999. **286**(5439): p. 509–512.
2. Albert, R., H. Jeong, and A.-L. Barabási, *Error and attack tolerance of complex networks.* Nature, 2000. **406**(6794): p. 378–382.
3. Buldyrev, S.V., et al., *Catastrophic cascade of failures in interdependent networks.* Nature, 2010. **464**(7291): p. 1025–1028.
4. Newman, M.E., *The structure of scientific collaboration networks.* Proceedings of the National Academy of Sciences, 2001. **98**(2): p. 404–409.
5. Jin, C., et al., *Emergence of scaling in complex substitutive systems.* Nature Human Behaviour, 2019. **3**(8): p. 837–846.
6. Watts, D.J., *A simple model of global cascades on random networks.* Proceedings of the National Academy of Sciences, 2002. **99**(9): p. 5766–5771.
7. Evans, J.A. and J.G. Foster, *Metaknowledge.* Science, 2011. **331**(6018): p. 721–725.
8. Fortunato, S., et al., *Science of science.* Science, 2018. **359**(6379): p. eaao0185.
9. Wang, D. and A.-L. Barabási, *The science of science*. 2021: Cambridge University Press.
10. Colizza, V., et al., *The role of the airline transportation network in the prediction and predictability of global epidemics.* Proceedings of the National Academy of Sciences, 2006. **103**(7): p. 2015–2020.
11. Brockmann, D. and D. Helbing, *The hidden geometry of complex, network-driven contagion phenomena.* Science, 2013. **342**(6164): p. 1337–1342.
12. Onnela, J.-P. and F. Reed-Tsochas, *Spontaneous emergence of social influence in online systems.* Proceedings of the National Academy of Sciences, 2010. **107**(43): p. 18375–18380.
13. Wang, B., et al., *Peer Review and the Diffusion of Ideas.* arXiv preprint arXiv:2507.11825, 2025.
14. Meng, F., et al., *Spreading dynamics of information on online social networks.* Proceedings of the National Academy of Sciences, 2025. **122**(4): p. e2410227122.
15. Kuhn, T., *The structure of scientific revolutions*. 1962: University of Chicago Press.
16. Saavedra, S., F. Reed-Tsochas, and B. Uzzi, *Asymmetric disassembly and robustness in declining networks.* Proceedings of the National Academy of Sciences, 2008. **105**(43): p. 16466–16471.
17. Wang, D., C. Song, and A.-L. Barabási, *Quantifying long-term scientific impact.* Science, 2013. **342**(6154): p. 127–132.
18. Yin, Y. and D. Wang, *The time dimension of science: Connecting the past to the future.* Journal of Informetrics, 2017. **11**(2): p. 608–621.
19. Candia, C., et al., *The universal decay of collective memory and attention.* Nature Human Behaviour, 2019. **3**(1): p. 82–91.
20. Lee, E.D., et al., *Synthesis of innovation and obsolescence.* arXiv preprint arXiv:2505.05182, 2025.
21. Haldane, A.G. and R.M. May, *Systemic risk in banking ecosystems.* Nature, 2011. **469**(7330): p. 351–355.
22. Dunne, J.A., R.J. Williams, and N.D. Martinez, *Network structure and biodiversity loss in food webs: robustness increases with connectance.* Ecology Letters, 2002. **5**(4): p. 558–567.

23. Gao, J., B. Barzel, and A.-L. Barabási, *Universal resilience patterns in complex networks.* Nature, 2016. **530**(7590): p. 307–312.
24. Artime, O., et al., *Robustness and resilience of complex networks.* Nature Reviews Physics, 2024. **6**(2): p. 114–131.
25. Hill, R., et al., *The pivot penalty in research.* Nature, 2025: p. 1–8.
26. Sinha, A., et al. *An overview of Microsoft Academic Service (MAS) and applications*. in *Proceedings of the 24th International Conference on World Wide Web*. 2015.
27. Lou, B. and L. Wu, *AI on Drugs: Can Artificial Intelligence Accelerate Drug Development? Evidence from a Large-scale Examination of Bio-pharma Firms.* MIS Quarterly, 2021. **45**(3).
28. Krieger, J.L., *Trials and terminations: Learning from competitors' R&D failures.* Management Science, 2021. **67**(9): p. 5525–5548.
29. Pastor-Satorras, R., et al., *Epidemic processes in complex networks.* Reviews of Modern Physics, 2015. **87**(3): p. 925.
30. Hethcote, H.W., *The mathematics of infectious diseases.* SIAM review, 2000. **42**(4): p. 599–653.
31. Scarpino, S.V. and G. Petri, *On the predictability of infectious disease outbreaks.* Nature Communications, 2019. **10**(1): p. 898.
32. Ruan, Z., et al., *Kinetics of social contagion.* Physical Review Letters, 2015. **115**(21): p. 218702.
33. Iacopini, I., et al., *Simplicial models of social contagion.* Nature Communications, 2019. **10**(1): p. 2485.
34. Hill, A.L., et al., *Infectious disease modeling of social contagion in networks.* PLOS Computational Biology, 2010. **6**(11): p. e1000968.
35. de Arruda, G.F., G. Petri, and Y. Moreno, *Social contagion models on hypergraphs.* Physical Review Research, 2020. **2**(2): p. 023032.
36. Karsai, M., et al., *Complex contagion process in spreading of online innovation.* Journal of The Royal Society Interface, 2014. **11**(101): p. 20140694.
37. Pastor-Satorras, R. and A. Vespignani, *Epidemic spreading in scale-free networks.* Physical Review Letters, 2001. **86**(14): p. 3200.
38. Johnson, S., J.J. Torres, and J. Marro, *Nonlinear preferential rewiring in fixed-size networks as a diffusion process.* Physical Review E—Statistical, Nonlinear, and Soft Matter Physics, 2009. **79**(5): p. 050104.
39. Brot, H., et al., *Edge removal balances preferential attachment and triad closing.* Physical Review E—Statistical, Nonlinear, and Soft Matter Physics, 2013. **88**(4): p. 042815.
40. Laird, S. and H.J. Jensen, *A steady state network model with a 1/k scale-free degree distribution.* arXiv preprint cond-mat/0603199, 2006.
41. Lim, S., et al., *Preferential detachment during human brain development: age-and sex-specific structural connectivity in diffusion tensor imaging (DTI) data.* Cerebral Cortex, 2015. **25**(6): p. 1477–1489.
42. Burns, L.R. and D.R. Wholey, *Adoption and abandonment of matrix management programs: Effects of organizational characteristics and interorganizational networks.* Academy of Management Journal, 1993. **36**(1): p. 106–138.
43. Greve, H.R., *Jumping ship: The diffusion of strategy abandonment.* Administrative Science Quarterly, 1995: p. 444–473.

44. Rao, H., H.R. Greve, and G.F. Davis, *Fool's gold: Social proof in the initiation and abandonment of coverage by Wall Street analysts.* Administrative Science Quarterly, 2001. **46**(3): p. 502–526.
45. Berger, J. and G. Le Mens, *How adoption speed affects the abandonment of cultural tastes.* Proceedings of the National Academy of Sciences, 2009. **106**(20): p. 8146–8150.
46. Gaba, V. and G. Dokko, *Learning to let go: Social influence, learning, and the abandonment of corporate venture capital practices.* Strategic Management Journal, 2016. **37**(8): p. 1558–1577.
47. Greenwood, B.N., et al., *The when and why of abandonment: The role of organizational differences in medical technology life cycles.* Management Science, 2017. **63**(9): p. 2948–2966.
48. Christakis, N.A. and J.H. Fowler, *The spread of obesity in a large social network over 32 years.* New England Journal of Medicine, 2007. **357**(4): p. 370–379.
49. Katz, M.L. and C. Shapiro, *Network externalities, competition, and compatibility.* The American Economic Review, 1985. **75**(3): p. 424–440.
50. Aral, S., L. Muchnik, and A. Sundararajan, *Distinguishing influence-based contagion from homophily-driven diffusion in dynamic networks.* Proceedings of the National Academy of Sciences, 2009. **106**(51): p. 21544–21549.
51. Aral, S. and D. Walker, *Identifying influential and susceptible members of social networks.* Science, 2012. **337**(6092): p. 337–341.
52. Schilling, M.A., *Technology success and failure in winner-take-all markets: The impact of learning orientation, timing, and network externalities.* Academy of Management Journal, 2002. **45**(2): p. 387–398.
53. Schumpeter, J.A., *Capitalism, Socialism, and Democracy*. 1942: Harper & Brothers.
54. White, H.C., *Chains of opportunity*. 2013: Harvard University Press.
55. Cohen, R., et al., *Resilience of the internet to random breakdowns.* Physical Review Letters, 2000. **85**(21): p. 4626.
56. Cohen, R., et al., *Breakdown of the internet under intentional attack.* Physical Review Letters, 2001. **86**(16): p. 3682.
57. Barabási, A.-L., *Network science*. 2016: Cambridge University Press.
58. Newman, M., *Networks*. 2018: Oxford University Press.
59. Callaway, D.S., et al., *Network robustness and fragility: Percolation on random graphs.* Physical Review Letters, 2000. **85**(25): p. 5468.
60. Li, M., et al., *Percolation on complex networks: Theory and application.* Physics Reports, 2021. **907**: p. 1–68.
61. Allard, A., et al., *Asymmetric percolation drives a double transition in sexual contact networks.* Proceedings of the National Academy of Sciences, 2017. **114**(34): p. 8969–8973.
62. Dodds, P.S. and D.J. Watts, *Universal behavior in a generalized model of contagion.* Physical Review Letters, 2004. **92**(21): p. 218701.
63. Garcia, D., P. Mavrodiev, and F. Schweitzer. *Social resilience in online communities: The autopsy of Friendster*. in *Proceedings of the first ACM Conference on Online Social Networks*. 2013.
64. Török, J. and J. Kertész, *Cascading collapse of online social networks.* Scientific Reports, 2017. **7**(1): p. 16743.

65. Yehezkel, A. and R. Cohen, *Degree-based attacks and defense strategies in complex networks.* Physical Review E, 2012. **86**(6): p. 066114.
66. Tian, L., et al., *Articulation points in complex networks.* Nature Communications, 2017. **8**(1): p. 1–9.
67. Motter, A.E. and Y.-C. Lai, *Cascade-based attacks on complex networks.* Physical Review E, 2002. **66**(6): p. 065102.
68. Goh, K.-I., et al., *Sandpile on scale-free networks.* Physical Review Letters, 2003. **91**(14): p. 148701.
69. Dobson, I., et al., *Complex systems analysis of series of blackouts: Cascading failure, critical points, and self-organization.* Chaos: An Interdisciplinary Journal of Nonlinear Science, 2007. **17**(2): p. 026103.
70. Su, Q., et al., *Evolution of prosocial behaviours in multilayer populations.* Nature Human Behaviour, 2022. **6**(3): p. 338–348.
71. Su, Q., A. McAvoy, and J.B. Plotkin, *Strategy evolution on dynamic networks.* Nature Computational Science, 2023. **3**(9): p. 763–776.
72. Chiba-Okabe, H. and J.B. Plotkin, *Social learning with complex contagion.* Proceedings of the National Academy of Sciences, 2024. **121**(49): p. e2414291121.

# Supplementary Information for
# Quantifying the Dynamics of Network Detachment Across Scientific, Technological, Commercial, and Pharmacological Domains

Binglu Wang[1,2,3*], Ching Jin[1,2,3,4*], Chaoming Song[5], Johannes Bjelland[6], Brian Uzzi[1,2,3] & Dashun Wang[1,2,3†]

[1] *Center for Science of Science and Innovation, 1800 Sherman Avenue, Evanston, IL 60201, USA*

[2] *Northwestern Institute on Complex Systems (NICO), 600 Foster Street, Evanston, IL 60208, USA*

[3] *Kellogg School of Management, 2211 Campus Drive, Evanston, IL 60208, USA*

[4] *Centre for Interdisciplinary Methodologies, University of Warwick, Coventry, CV4 7AL, United Kingdom*

[5] *Department of Physics, University of Miami, 1320 S Dixie Hwy, Coral Gables, FL 33146, USA*

[6] *Telenor Research and Development, Snarøyveien 30 N-1360 Fornebu, Norway*

**These authors contributed equally to this work.*

[†] *Correspondence should be addressed to D.W. (dashun.wang@kellogg.northwestern.edu)*

# Contents

# S1 Dataset Descriptions

## S1.1 $D_1$: Detachment in Scientific Fields

The first dataset ($D_1$) tracks the detachment dynamics of scientific fields through the publication records of individual scientists. The data are drawn from the Microsoft Academic Graph (MAG), which contains approximately 172 million publications and their associated metadata[1]. We use the MAG field classification system to identify research domains[2].

*Identifying detachment from scientific fields for each scientist:* MAG organizes research fields in a hierarchical taxonomy. We focus on leaf-level fields (i.e., fields without child fields). When a field contains fewer than 500 publications or has more than three hierarchical parent levels, we instead use the corresponding parent fields to ensure sufficient observations and stable classifications. To determine when a scientist detaches from a field, we identify the year in which the scientist's primary research focus shifts away from that field after having worked in it for at least two consecutive years (Fig. S1). Specifically, we implement the following procedure: (1) compile all publications associated with each scientist that are linked to at least one leaf field; (2) for each year, identify the field in which the scientist published the largest number of papers and define it as the scientist's primary research focus for that year; and (3) construct the scientist's career trajectory based on the sequence of their annual primary research focuses. A detachment event occurs when a field no longer appears as the scientist's primary research focus in subsequent years. Our analysis includes scientists whose publishing careers span at least 10 years and who have published at least 10 papers.

Overall, $D_1$ includes 990K scientists working across 1,219 scientific fields that emerged after 1940 and were studied by at least 500 scientists from 1940 to 2015.

## S1.2 $D_2$: Detachment in Patenting

The second dataset ($D_2$) captures detachment dynamics in technological domains using 6.9M patents granted by the United States Patent and Trademark Office (USPTO) from 1976 to 2015. Technological domains are approximated using the Cooperative Patent Classification (CPC) system, focusing on CPC subgroup codes.

*Identifying detachment from patenting domains for each inventor:* Following a procedure similar to that used for $D_1$, we reconstruct each inventor's annual patenting focus. For each year, we identify the CPC subgroup in which the inventor produces the largest number of patents and define this subgroup as the inventor's primary patenting domain for that year. Based on these annual domain assignments, a detachment event occurs when an inventor ceases to focus their patenting activity on a previously dominant domain.

To ensure adequate coverage, we restrict the analysis to patenting domains involving at least 200 inventors during the observation period. The resulting dataset includes 123K inventors with patenting careers spanning more than five years across 291 technological domains from 1976 to

2015.

## S1.3 $D_3$: Detachment from Mobile Handsets

The third dataset ($D_3$) captures the commercial domain of mobile handsets, recording individual usage history provided by a European telecommunication company. We identify detachment from mobile handsets by tracking changes in their daily active users from 2006 to 2014.

*Identifying detachment from mobile handsets for each consumer:* We determine the detachment from mobile handsets by each user through SIM card records, which provide clear timelines for when users change mobile phones. Specifically, we use the last usage day of a mobile handset as the user's detachment time from that handset.

We focus on mobile handsets used by more than 1,000 users. In total, $D_3$ records 3.5M users using 704 types of mobile handsets from 2006 to 2014.

For the network-based analyses, we approximate the communication network among mobile-phone users using anonymized call-detail records over a three-month window from October 1, 2013, to January 1, 2014. Two users are connected by an edge if they exchanged at least one phone call during this window. The network-based detachment analyses are conducted on the subset of 1.7M users observed in these call records, for whom both handset usage and communication ties can be measured. Users without communication-network observations are retained in the aggregate handset life-cycle analyses but are excluded from analyses requiring network-neighbor information.

## S1.4 $D_4$: Detachment in Pharmacological Innovation

The fourth dataset ($D_4$) examines detachment dynamics in pharmaceutical innovation using the Cortellis Drug Discovery Intelligence database. This database contains detailed information on global drug development pipelines and covers 648,702 drug candidates over the past 25 years.

*Identifying detachment from mechanisms of action (MOAs) for each pharmaceutical organization:* We characterize pharmacological innovation using biological mechanisms of action (MOAs) associated with drug candidates. For each pharmaceutical company, we track its portfolio of MOAs used in clinical development. A detachment event occurs when a pharmaceutical organization ceases to pursue a previously adopted MOA in clinical trials.

We focus on second-level or higher MOA categories adopted by at least 10 pharmaceutical organizations. The resulting dataset tracks 248 MOAs used by 5,133 pharmaceutical companies and measures the fraction of organizations pursuing each MOA in drug development.

# S2 Empirical Methods

## S2.1 Network Structure of Different Social Systems

The present study investigates the decline dynamics of robust network systems, which represent a predominant class of networks in social contexts[3]. As a first step, we examine the structural properties of the four empirical systems analyzed in this study (i.e., the scientific field dataset, patenting dataset, mobile handset dataset, and drug development dataset). Specifically, we plot the degree distributions of the corresponding social networks for each dataset (Fig. S2).

Across all four systems, we observe that the degree distributions exhibit heavy-tailed patterns, consistent with the heterogeneous connectivity commonly observed in real-world social networks. These findings confirm that the empirical systems examined in this study belong to the class of heterogeneous networks that have been widely studied in the literature on network science.

To further characterize the structural nature of the empirical item-specific networks, we calculate summary statistics for the networks used in the preferential-detachment analyses. For each domain, we report the number of items, the average size of the corresponding item-specific networks, and the estimated power-law exponent of the item-level degree distributions in Table S1. The corresponding degree distributions are shown in Fig. S2. Together, these results show that all four empirical settings exhibit strongly heterogeneous item-level network structures, with estimated power-law exponents ranging from approximately 1.8 to 3.0.

Table S1: **Structural statistics of the empirical item-specific networks.** For each domain, we construct item-specific relational networks for the focal scientific field, patenting domain, handset model, or drug MOA retained in the main life-cycle analysis. The table reports the number of items, the average size of the corresponding item-level networks, and the estimated power-law exponent $\gamma$ of the item-level degree distributions.

| Domain | Number of items | Average size of item network | Power-law exponent $\gamma$ |
|---|---|---|---|
| $D_1$ Science fields | 1,219 | 1,078 | 2.99 |
| $D_2$ Patents | 291 | 500 | 2.98 |
| $D_3$ Handsets | 704 | 4,926 | 2.97 |
| $D_4$ Drug MOAs | 248 | 96 | 1.79 |

## S2.2 Selecting Time Intervals to Measure the Structural Effect

In Fig. 2 of the main text, we measure the structural effect within a short time interval to control for potential temporal confounding effects. To ensure comparability across the four empirical domains, we focus on the stage of the life cycle when an innovation reaches its half-life, defined as the time at which its popularity has declined to half of its peak level.

For datasets $D_1$, $D_2$, and $D_4$, we measure the structural effect over the minimal time resolution available in the data, around the half-life. For dataset $D_3$, which records daily user activity, we measure detachment events within a 100-day window around the half-life.

## S2.3 Robustness Check Across Different Time Intervals

To examine the robustness of the observed structural effects, we repeat the analyses presented in Fig. 2 of the main text using alternative points along the life-cycle dynamics. In addition to the half-life (50% of peak popularity), we conduct the same analyses when the popularity has declined to 60% (Fig. S3) and 40% (Fig. S4) of its maximum value. The results indicate that the observed patterns remain consistent across these alternative time intervals.

In both Fig. S3 and Fig. S4, we divide individuals and organizations into two groups based on whether detachment events have occurred among their network neighbors ($r_i > 0$ versus $r_i = 0$), where $r_i$ denotes the number of neighbors of node $i$ that have already detached. We then measure the detachment rate separately for the two groups. The distribution of detachment rates shows visibly different patterns (panels A–D in both figures), with the distribution for the group $r_i > 0$ systematically shifting to the right relative to the $r_i = 0$ group. This pattern indicates that the detachment rate tends to be higher when detachment events have occurred among an individual's network neighbors.

To systematically evaluate this relationship across innovations in different domains, we estimate the relationship between detachment probability $\nu_i$ and the number of detached neighbors $r_i$ for each scientific field, patenting domain, mobile handset, and drug MOA. Specifically, we fit a linear model for each innovation and examine the distribution of the estimated slope coefficients. We find that a substantial fraction of innovations across the four domains exhibit a positive slope, indicating that detachment probability increases with the number of neighbors who have already detached. In particular, the fraction of innovations with positive slopes equals 73%, 72%, 79%, and 64% for scientific fields, patenting domains, mobile handsets, and drug MOAs, respectively, when popularity declines to 60% (Fig. S3E–H), and 71%, 57%, 78%, and 60%, respectively, when popularity declines to 40% (Fig. S4E–H).

Together, these results confirm that the structural pattern of preferential detachment remains robust across different points along the life-cycle dynamics.

## S2.4 Robustness Check on Averaged Detachment Curves

In addition to the representative item-level examples shown in Fig. 2E–H of the main text, we also calculate averaged detachment curves for each domain. This provides a population-level visualization of the relationship between detachment probability and the number of detached neighbors. As shown in Fig. S5, the averaged curves are consistent with the main empirical pattern reported in Fig. 2.

## S2.5 Robustness Check on Normalized Exposure, $r_i/k_i$

To test whether the observed preferential-detachment pattern is driven only by the raw number of detached neighbors, we repeat the main empirical analysis using normalized exposure, $r_i/k_i$, where $r_i$ is the number of detached neighbors of node $i$, and $k_i$ is its degree in the item-specific relational network. This analysis asks whether detachment probability increases with the fraction of a node's neighbors who have already detached. As shown in Fig. S6, the main pattern remains consistent across the four empirical domains, confirming that the observed preferential-detachment pattern is not solely driven by the raw number of detached neighbors.

## S2.6 Robustness Check on $\alpha$ Distribution

To connect the empirical detachment patterns to the model parameter $\alpha$, we estimate an item-level empirical analogue of preferential-detachment strength from the fitted relationship between detachment probability and the number of detached neighbors. Positive fitted slopes correspond to $\alpha > 0$, consistent with preferential rather than random detachment. Figure S7 reports the estimated domain-level values of $\alpha$, and Table S2 provides the corresponding summary statistics. These results show that the estimated preferential-detachment strength is positive across all four empirical domains.

Table S2: **Summary statistics for empirical preferential-detachment strength.** The table reports, for each empirical domain, the number of items analyzed, the mean slope of the relationship between $\nu_i$ and $r_i$, the percentage of items with positive slopes, and the estimated item-level preferential detachment parameter $\alpha$. Positive slopes correspond to $\alpha > 0$, consistent with preferential rather than random detachment. Across all four domains, most items exhibit positive slopes, supporting the generality of preferential detachment across scientific, technological, commercial, and pharmacological systems. The table only includes items with a well-defined half-life observation window and within-item variation in exposure. The resulting item counts are therefore a subset of the full item sets reported in Table S1.

| Domain | Number of items | Mean slope | Positive slope (%) | Mean item-level $\alpha$ |
|---|---|---|---|---|
| $D_1$ Science fields | 508 | 0.0349 | 73 | 0.1657 |
| $D_2$ Patents | 125 | 0.0411 | 70 | 0.1410 |
| $D_3$ Handsets | 513 | 0.0335 | 79 | 0.1235 |
| $D_4$ Drug MOAs | 73 | 0.1390 | 70 | 0.3153 |

# S3 Heterogeneous Mean-Field Theory

## S3.1 Method Description

To better understand the phase transition phenomenon in the small $\alpha$ region, we adopt the heterogeneous mean-field approximation (HMA)—a frequently used analytical method in epidemiology, network science, and percolation theory[4,5]. HMA assumes that nodes with the same degree follow the same behavior. By denoting $i_k$ as the survival probability of a node with degree $k$, we can express the fraction of detached nodes $f$ in the system as:

$$f = 1 - \sum_k P_k i_k, \tag{S1}$$

where $P_k$ is the degree distribution.

To identify the critical point $f_c$ at which the giant component vanishes, we calculate $u$, the probability that a remaining node does not belong to the giant component. $u$ is a function of $f$ satisfying the following equation[5]:

$$u = \sum_k q_k \left(1 - i'_{k+1} + i'_{k+1} u^k\right), \tag{S2}$$

where $q_k$ corresponds to the excess degree distribution of the network, and $i'_{k+1}$ is the *neighbor-weighted survival probability* for nodes with degree $k+1$. This probability quantifies the survival probability of a node with degree $k+1$ when sampled through surviving continuation edges, such that configurations with more surviving continuation neighbors receive proportionally greater weight. Notice that $i'_{k+1}$ is different from $i_{k+1}$. $i_{k+1}$ is a *direct survival probability*—it directly counts how many nodes with degree $k+1$ survive, without considering how these nodes are connected to others. $i'_{k+1}$ is more complex. Here, we consider the process of reaching nodes through their neighbors, which introduces a bias: nodes with more connections (i.e., higher-degree nodes) are more likely to be picked multiple times since they have more neighbors. We present the quantitative relationship between $i_{k+1}$ and $i'_{k+1}$ in later sections.

Equation (S2) suggests that if a node is not connected to the giant component, it either links to a node that is also not in the giant component or the node has already been removed. Based on percolation theory, the giant component vanishes when the derivative of the right-hand side equals 1 at $u = 1$. In other words:

$$\left[\frac{d}{du}\sum_k q_k\left(1 - i'_{k+1} + i'_{k+1}u^k\right)\right]_{u=1} = 1. \tag{S3}$$

The phase transition point $f_c$ can be obtained through equation (S3). To achieve this, we need to express $i_k$ in terms of $f$ and substitute it back into equation (S3).

## S3.2 Express $i_k$ as a Function of $t$

Let us first express $i_k$ in terms of $t$. We have:

$$\frac{di_k}{dt} = -i_k\nu_0[\alpha k\mu(t) + (1-\alpha)]. \tag{S4}$$

Here, $\nu_0$ is an overall time-scale factor in the continuous-time hazard formulation; because the process in the main-text simulations is implemented as a normalized selection probability over surviving nodes, $\nu_0$ rescales time but does not affect the order of node removals or the critical threshold $f_c$. $\mu(t) \equiv 1 - \sum_{k'} q_{k'}(i'_{k'+1})$ measures the probability that a remaining node is linked to a neighbor that has detached. Solving equation (S4), we have:

$$i_k = e^{-\nu_0(1-\alpha)t}[h(t)]^k, \tag{S5}$$

where $h$ is a function satisfying:

$$\frac{dh}{dt} = -\nu_0 \alpha \mu h. \tag{S6}$$

By inserting equation (S5) into the definition of $\mu$, we can express $\mu$ as a function of $h$:

$$\mu(t) = 1 - e^{-\nu_0(1-\alpha)t} G_1(h)/h, \tag{S7}$$

where $G_1(x) \equiv \sum_k \frac{1}{<k>}(k+1)P(k+1)x^k$ is the generating function for excess degree. Here we have also used an important relation: $i'_k/i_k = 1/h^2$, which we prove in S3.3. Combining equations (S7) and (S6), we have:

$$\frac{dh}{dt} = -\nu_0 \alpha (h - e^{-\nu_0(1-\alpha)t} G_1(h)). \tag{S8}$$

## S3.3 Relationship Between $i'_k$ and $i_k$

Since $i'_k$ measures the survival probability of k-degree nodes through the nodes' surviving neighbors, nodes with more surviving neighbors will contribute more to the probability. We have:

$$\begin{aligned}
\frac{di'_k}{dt} &= -i'_k \nu_0 \left[ (1-\alpha) + \alpha \frac{\sum_{m=0}^{k-1} (k-1-m) m \binom{k-1}{m} (1-\mu)^m \mu^{k-m-1}}{\sum_{m=0}^{k-1} m \binom{k-1}{m} (1-\mu)^m \mu^{k-m-1}} \right] \\
&= -i'_k \nu_0 \left[ (1-\alpha) + \alpha \frac{(k-1)^2(1-\mu) - (k-1)(1-\mu)\mu - (k-1)^2(1-\mu)^2}{(k-1)(1-\mu)} \right] \\
&= -i'_k \nu_0 \left[ (1-\alpha) + \alpha (k-2)\mu \right].
\end{aligned} \tag{S9}$$

where $m$ measures the number of surviving neighbors of the studied node among the $k-1$ neighbors other than the incoming surviving neighbor. Therefore, $k-1-m$ is the number of removed neighbors. The reason the sum runs from 0 to $k-1$ instead of to $k$ is that the node is reached through a surviving node. Solving this equation, we obtain:

$$i'_k = e^{-\nu_0(1-\alpha)t}[h(t)]^{k-2}. \tag{S10}$$

Combining equations ( S10) and (S5), we have:

$$i'_k / i_k = \frac{1}{h^2} \tag{S11}$$

## S3.4 Express $i_k$ as a Function of $f$

Combining equations ( S5) and (S1), we have:

$$e^{-\nu_0(1-\alpha)t} = \frac{1-f}{G_0(h)}. \tag{S12}$$

By inserting this result back into equations ( S5) and (S8), we have:

$$i_k = \frac{1-f}{G_0(h)} h^k, \tag{S13}$$

and

$$\frac{dh}{df} = \frac{-\alpha h G_0^2(h) + \alpha G_0(h) G_1(h)(1-f)}{\alpha(1-f) h G_0(h) G'_0(h) - \alpha(1-f)^2 G'_0(h) G_1(h) + (1-\alpha)(1-f) G_0^2(h)}, \tag{S14}$$

where $G_0$ and $G_1$ are generating functions, defined as $G_0(x) \equiv \sum_k P(k)x^k$, $G_1(x) \equiv \sum_k \frac{1}{<k>}(k+1)P(k+1)x^k$. Inserting equation (S13) into equation (S3), we obtain the self-consistent equation

for $f_c$:

$$f_c = 1 - \frac{G_0(h_c) < k >}{G_0''(h_c)} = 1 - \frac{< k > \sum_k P_k h_c^k}{\sum_k P_k (k^2 - k) h_c^{k-2}}, \tag{S15}$$

where $h_c$ is the solution of equation ( S14) for $f = f_c$.

## S3.5 Molloy–Reed Criterion Method

An alternative method for identifying the phase transition point is through the Molloy–Reed criterion[3,6]: a network with any degree distribution has a giant component if

$$\frac{< k^2 >}{< k >} > 2. \tag{S16}$$

We first calculate the average degree of a network when a fraction $f$ of the nodes has been removed:

$$\begin{aligned} < k' >_f &= \frac{\sum P_k i_k \binom{k}{k'} k' \mu^{k-k'} (1-\mu)^{k'}}{\sum P_k i_k} \\ &= \frac{\sum_k k h^k P_k (1-\mu)}{G_0(h)} \\ &= \frac{(1-\mu) G_0'(h) h}{G_0(h)} \\ &= \frac{(1-f) G_0'(h) G_1(h)}{G_0^2(h)}. \end{aligned} \tag{S17}$$

In the same way, we can calculate $< k'^2 - k' >$:

$$\begin{aligned} < k'^2 - k' >_f &= \frac{\sum P_k i_k \binom{k}{k'} (k'^2 - k') \mu^{k-k'} (1-\mu)^{k'}}{\sum P_k i_k} \\ &= \frac{\sum_k (k^2 - k) h^k P_k (1-\mu)^2}{G_0(h)} \\ &= (1-f)^2 \frac{G_0''(h) G_1^2(h)}{G_0^3(h)}. \end{aligned} \tag{S18}$$

By combining these results, we have:

$$f_c = 1 - \frac{G_0(h_c) < k >}{G_0''(h_c)} = 1 - \frac{< k > \sum_k P_k h_c^k}{\sum_k P_k (k^2 - k) h_c^{k-2}}, \tag{S19}$$

which is consistent with our previous findings.

# S4 Relationship Between $k_f$ and $r$

## S4.1 Relationship Between $k_f$ and $r$ in Different Network Structures

Why is the relationship between $k_f$ and $r$ different across various networks? An intuitive understanding can be gained by analyzing the initial step of the process, where the first fraction $f_0$ of nodes is removed from a generated network. The process is a simple Poisson process, where the joint probability of the current degree $k$ and $r$ can be written as: $P_{f_0}(k, r) = P(k + r)\binom{k+r}{r}(1 - f)^k f^r$. For an ER network, the formation process is also a Poisson process, where each node is independently connected to other nodes with the same probability. Since each node behaves independently in both the formation and removal processes, the distributions of $k$ and $r$ should also be independent. Mathematically, we have:

$$P_{f_0}(k, r) = \frac{\lambda^{k+r} e^{-\lambda}}{(k + r)!} \binom{k + r}{r} (1 - f)^k f^r = \frac{[\lambda(1 - f)]^k e^{-\lambda(1-f)}}{k!} \frac{[\lambda f]^r e^{-\lambda f}}{r!} \tag{S20}$$

Here, $P_{f_0}(k, r)$ can be decoupled into two independent distributions for $k$ and $r$, suggesting that the process can be viewed as two independent processes. In the next section, we show that not only in the initial step but in all steps of the ER network, the joint probability can be decoupled, demonstrating the independence of the current degree $k$ and $r$. The key to this decoupling property is the factorial decay of the Poisson process, which must satisfy two criteria: 1) each node is independently attached to other nodes (independence), and 2) each node attaches to other nodes with the same probability (homogeneity).

For a network with heavy tails, the formation process can be represented as an inhomogeneous Poisson process (see hidden variable network models), where each node attaches to others independently but with a different probability. A node with a higher probability is more likely to have a higher $r$ and a higher current degree, presenting a positive correlation between $k$ and $r$. In contrast, for a network with narrower tails, nodes do not link to others independently, and the degree for each node is bounded. Therefore, removing more neighbors reduces the current degree.

To further understand the process, we directly study the relationship between $k_f$ and $r$ for different $f$ (Fig. 3C in the main text). Interestingly, we find that the curves are parallel to each other when $r$ is relatively large, suggesting that the $k_f - r$ curve is determined only by the network topology in that condition, and it is independent of the specific dynamics. To test this hypothesis, and to systematically uncover the role of network topology in the detachment process, we quantify the relationship between $k_f$ and $r$ for networks with varying levels of degree heterogeneity. We find that although the relationship between $k_f$ and $r$ is complex (see the next section for the full mathematical derivation), the asymptotic behavior of $k_f(r)$ is very simple:

$$k_f(r)\Big|_{r\to\infty} = W(h)H(r) \sim H(r) \tag{S21}$$

It can be decoupled into two independent factors, $H(r) \equiv \frac{P_k(r+1)}{P_k(r)}(r+1)$, and $P_k(x)$ takes the form of the degree distribution of the network. $H$ quantifies how heavy the degree distribution is compared to a Poisson distribution, and

$$W(h) = \begin{cases} \frac{(1-f)G_1(h)}{G_0(h)-(1-f)G_1(h)} & \text{if } H(r)|_{r\to\infty} \to \infty, \\ (1-f)G_1(h)/G_0(h) & \text{if } H(r)|_{r\to\infty} = H_0. \end{cases} \tag{S22}$$

Since $W(h)$ is a function independent of $r$, equation (S21) suggests that the trend of $k_f(r)$ is solely determined by $H$, which depends only on the degree distribution of the network. The second branch includes finite limiting values of $H_0$, including $H_0 = 0$.

For ER networks, since the degree distribution follows a Poisson distribution (Fig. S8E), we can calculate that $\frac{P_k(x+1)}{P_k(x)} = \frac{\lambda}{x+1}$, and $H(x) = \lambda$, suggesting that $k_f(r)$ remains constant when $r$ is large (Fig. 3F in the main text). Therefore, under preferential detachment, where nodes with higher $r$ are more likely to be detached, we effectively select nodes with a random degree $k_f$, supporting the conclusion that preferential detachment is identical to random detachment on ER networks. Importantly, for ER networks, $k_f(r)$ remains constant not only for large $r$ but for any given $r$. This result corroborates our findings in Fig. S8F, explaining why the phase transition point does not change in an ER network. It also echoes the analytical results demonstrated in equation (2) of the main text. The generating functions for ER random networks take special forms: $G_0(x) = G_1(x) = e^{\langle k\rangle(x-1)}$. By substituting this back into equation (2) of the main text, we recover the Molloy–Reed criterion:

$$f_c = 1 - \frac{\langle k \rangle}{\langle k^2 \rangle - \langle k \rangle},$$

which is independent of $\alpha$ (Fig. 3E in the main text, solid line). This result validates that preferential detachment is exactly the same dynamic process as random detachment on ER networks.

Empirical networks typically exhibit heavy-tailed degree distributions. From equation (4) in the main text, we find that if the degree distribution has a heavier tail than the Poisson distribution, such as a power-law distribution (Fig. S8B), $H(x)$ grows faster than a constant, indicating a positive correlation between $k_f$ and $r$ (Fig. 3C in the main text). Therefore, for these networks, if nodes are removed under the preferential detachment process, hubs are more likely to be removed, leading to an earlier phase transition (Fig. S8C). This result supports the argument that preferential detachment behaves similarly to network attacks (red triangles in Fig. S8C) on heavy-tailed networks. As for networks with narrow degree distributions (i.e., narrower than the Poisson distribution), such as networks with a normal degree distribution (Fig. S8H), $H(x)$ decreases with $x$, suggesting a negative correlation between $k_f$ and $r$ (Fig. 3I in the main text). Interestingly, the negative relationship between $k_f$ and $r$ indicates that if we remove nodes with higher $r$ under preferential detachment, we are actually targeting nodes with lower remaining degrees, thus inducing a delayed percolation compared to random detachment (Fig. S8I).

Table S3 summarizes the predictions for several common networks:

Table S3: Summary of different network structures

| **Network** | **Degree Distribution** | $H(r)$ | $H(r)$ **(**$r \to \infty$**)** |
|---|---|---|---|
| Scale-free networks | $A(\gamma)k^{-\gamma}$ | $(r+1)(1+1/r)^{-\gamma}$ | $r$ |
| ER networks | $\frac{\lambda^k e^{-\lambda}}{k!}$ | $\lambda$ | $\lambda$ |
| Gaussian networks | $A(\mu_0,\sigma)e^{-\frac{1}{2}\left(\frac{k-\mu_0}{\sigma}\right)^2}$ | $(r+1)e^{\frac{-2r+2\mu_0-1}{2\sigma^2}}$ | 0 |

*Note: $A(\gamma)$ and $A(\mu_0,\sigma)$ are normalization constants. Here $\gamma$ denotes the power-law exponent, $\lambda$ denotes the mean degree of the Poisson distribution, and $\mu_0$ denotes the mean degree of the Gaussian degree distribution.*

## S4.2 $k_f$ as a Function of $r$

In this section, we discuss the average remaining degree of a node after removing a fraction $f$ of nodes, denoted by $k_f$, as a function of the number of removed neighbors $r$. We begin our analysis with the random detachment process ($\alpha = 0$) and then proceed to understand the behavior under preferential detachment.

**1. Random Detachment (**$\alpha = 0$**)**

For random detachment, we have the following expression:

$$\begin{aligned} k_f(r) &= \frac{\sum_{k'=0}^{\infty} k' p(k'+r)\binom{k'+r}{k'} f^r (1-f)^{k'}}{\sum_{k'=0}^{\infty} p(k'+r)\binom{k'+r}{k'} f^r (1-f)^{k'}} \\ &= \frac{G_r^{(r+1)}(1-f)}{G_r^{(r)}(1-f)}(1-f), \end{aligned} \tag{S23}$$

where $G_r(x) = \sum_{k=0}^{\infty} p(k+r)x^{k+r}$, $p(k)$ denotes the degree distribution of the network, and $G_r^{(n)}(x)$ represents the $n^{\text{th}}$ derivative of $G_r(x)$, i.e., $G_r^{(n)}(x) = \frac{\partial^n G_r(x)}{\partial x^n}$.

Although $G_r(x)$ is intricate and topology-dependent, its asymptotic behavior can be readily determined. In the limit $r \to \infty$, we have:

$$G_r^{(r)}(x) = \sum_{a=0}^{\infty} (r+a)!\, p(r+a) \frac{x^a}{a!}. \tag{S24}$$

Additionally, we can write:

$$\begin{aligned} G_r^{(r+1)}(x) &= \sum_{a=0}^{\infty} \frac{(r+a)!\, a}{a!} p(r+a) x^{a-1} \\ &= \sum_{a=0}^{\infty} \frac{(r+a+1)!}{a!} p(r+a+1) x^a. \end{aligned} \tag{S25}$$

Now, by combining equations ( S24) and (S25), we obtain:

$$\begin{aligned} (1-x)G_r^{(r+1)}(x) &= \sum_{a=0}^{\infty} \left[ \frac{(r+a+1)!\, p(r+a+1)}{a!} - \frac{(r+a)!\, a\, p(r+a)}{a!} \right] x^a \\ \Rightarrow \frac{(1-x)G_r^{(r+1)}(x)}{G_r^{(r)}(x)} &= \frac{\sum_{a=0}^{\infty} \left[ (r+a)!\, p(r+a) \frac{x^a}{a!} \right] \left[ \frac{(r+a+1)!}{(r+a)!} \frac{p(r+a+1)}{p(r+a)} - a \right]}{\sum_{a=0}^{\infty} (r+a)!\, p(r+a) \frac{x^a}{a!}}. \end{aligned} \tag{S26}$$

Let us define $H(r) \equiv (r+1)\frac{p(r+1)}{p(r)}$. Thus, we have:

$$k_f(r) = \frac{1-f}{f} \frac{\sum_{a=0}^{\infty} \left[ (r+a)!\, p(r+a) \frac{(1-f)^a}{a!} \right] \left[ H(r+a) - a \right]}{\sum_{a=0}^{\infty} (r+a)!\, p(r+a) \frac{(1-f)^a}{a!}}. \tag{S27}$$

Next, we consider the asymptotic properties of $k_f(r)$ for networks with varying forms of $H(r)$.

**Class 1:** $H(r)|_{r\to\infty} \to \infty$

For a network's degree distribution satisfying this condition, $H(r+a)-a \approx H(r)$ as $r \to \infty$. Substituting this into equation ( S27), we obtain:

$$k_f(r)|_{r\to\infty} = \frac{1-f}{f} H(r). \tag{S28}$$

**Example Class 1: Scale-Free Network**

A common example of this class is the scale-free network with a power-law degree distribution:

$$\begin{aligned} p(r) \sim r^{-\gamma} \quad \Rightarrow \quad H(r)|_{r\to\infty} &= (r+1)\left(1+\frac{1}{r}\right)^{-\gamma} \approx (r+1)\left(1-\frac{\gamma}{r}\right) \\ &= r - \gamma + 1 - \frac{\gamma}{r} = r + O(1) \to \infty. \end{aligned} \tag{S29}$$

Thus, for a scale-free network, we have:

$$k_f(r)|_{r\to\infty} = \frac{1-f}{f} H(r)|_{r\to\infty} = \frac{1-f}{f} r. \tag{S30}$$

**Class 2:** $H(r)|_{r\to\infty} = H_0$ **(**$H_0 > 0$**)**

For networks with a degree distribution satisfying this condition, we can directly derive:

$$
\begin{aligned}
k_f(r)|_{r\to\infty} &= \frac{1-f}{f}\frac{\sum_{a=0}^{\infty}\left[(r+a)!\,p(r+a)\frac{(1-f)^a}{a!}\right][H(r+a)-a]}{\sum_{a=0}^{\infty}(r+a)!\,p(r+a)\frac{(1-f)^a}{a!}} \\
&= \frac{1-f}{f}(H_0 - \frac{\sum_{a=0}^{\infty}(r+a)!\,p(r+a)\frac{(1-f)^a}{a!}(a)}{\sum_{a=0}^{\infty}(r+a)!\,p(r+a)\frac{(1-f)^a}{a!}}) \\
&= \frac{1-f}{f}(H_0 - \frac{\sum_{a=0}^{\infty}H_0^{r+a}\frac{(1-f)^a}{a!}(a)}{\sum_{a=0}^{\infty}H_0^{r+a}\frac{(1-f)^a}{a!}}) \\
&= \frac{1-f}{f}(H_0 - \frac{\sum_{a=0}^{\infty}H_0^{r+a}\frac{(1-f)^a}{a!}a}{\sum_{a=0}^{\infty}H_0^{r+a}\frac{(1-f)^a}{a!}}) \\
&= \frac{1-f}{f}(H_0 - \frac{\sum_{a=0}^{\infty}H_0^{a}\frac{(1-f)^a}{a!}a}{\sum_{a=0}^{\infty}H_0^{a}\frac{(1-f)^a}{a!}}) \\
&= \frac{1-f}{f}(H_0 - \frac{e^{H_0(1-f)}H_0(1-f)}{e^{H_0(1-f)}}) \\
&= (1-f)H_0 = (1-f)H(r)
\end{aligned}
\tag{S31}
$$

**Example Class 2: Erdős–Rényi (ER) Network**

For the Erdős–Rényi (ER) network, where the degree distribution follows a Poisson distribution with mean degree $\lambda$, $H(r)$ is constant:

$$
p(r) \sim \frac{e^{-\lambda}\lambda^r}{r!} \quad \Rightarrow \quad H(r) = (r+1)\frac{\lambda}{r+1} = H_0 = \lambda. \tag{S32}
$$

Since $H(r)$ remains constant for all $r$, we can calculate $G_r(x)$ for any $r$:

$$
\begin{aligned}
G_r(x) &= \sum_{k=0}^{\infty} p(k+r)x^{k+r} \\
&\sim \sum_{k=0}^{\infty}\frac{e^{-\lambda}\lambda^{k+r}}{(k+r)!}x^{k+r} \\
&= e^{-\lambda}\left(e^{\lambda x} - \sum_{n=0}^{r-1}\frac{(\lambda x)^n}{n!}\right).
\end{aligned}
\tag{S33}
$$

Taking the $r^{\text{th}}$ and $(r+1)^{\text{th}}$ derivatives of the second term results in its vanishing. Therefore, we have:

$$k_f(r) = \lambda(1-f) = (1-f)H_0, \tag{S34}$$

which is consistent with equation ( S31).

**Class 3:** $H(r)|_{r\to\infty} = 0$

For networks satisfying this condition, we revisit the definitions of $G_r^{(r)}$ and $G_r^{(r+1)}$. For

$$G_r^{(r)}(x) = \sum_{a=0}^{\infty}(r+a)!\,p(r+a)\frac{x^a}{a!}, \tag{S35}$$

we analyze the ratio of consecutive terms, yielding:

$$\frac{(r+a+1)!\,p(r+a+1)\frac{x^{a+1}}{(a+1)!}}{(r+a)!\,p(r+a)\frac{x^a}{a!}}|_{r\to\infty} = \frac{x}{1+a}H(r)|_{r\to\infty} = 0. \tag{S36}$$

Thus, only the first term dominates. The same applies to $G_r^{(r+1)}$, leading to:

$$G_r^{(r)}(x)|_{r\to\infty} = r!p(r), \tag{S37}$$

and

$$G_r^{(r+1)}(x)|_{r\to\infty} = (r+1)!p(r+1). \tag{S38}$$

By substituting these expressions into the definition of $k_f(r)$, we find:

$$k_f(r)|_{r\to\infty} = (1-f)(r+1)\frac{p(r+1)}{p(r)} = (1-f)H(r). \tag{S39}$$

**Example Class 3: Normal Degree Distribution**

For a network with a normal degree distribution:

$$p(k) \sim e^{-\frac{1}{2}\left(\frac{k-\mu_0}{\sigma}\right)^2}, \tag{S40}$$

we find:

$$H(r) = (r+1) \cdot \exp\left(\frac{-2r + 2\mu_0 - 1}{2\sigma^2}\right),$$

and therefore,

$$k_f(r)|_{r\to\infty} = (1-f)H(r)|_{r\to\infty} = 0. \tag{S41}$$

**Summary for Random Detachment**

In summary, for random detachment, the asymptotic behavior of $k_f(r)$ across the different classes can be expressed as:

$$k_f(r)\Big|_{r\to\infty} = L(f)H(r) \sim H(r), \tag{S42}$$

where $H(r) \equiv \frac{P_k(r+1)}{P_k(r)}(r+1)$ and

$$L(f) = \begin{cases} (1-f)/f & \text{if } H(r)|_{r\to\infty} \to \infty, \\ (1-f) & \text{if } H(r)|_{r\to\infty} = H_0. \end{cases} \tag{S43}$$

**2. Preferential Detachment ($\alpha > 0$)**

Next, we consider $k_f$ as a function of $r$ under preferential detachment. We have

$$k_f(r) = \frac{\sum_{k'=0}^{\infty} p(k'+r) i_{k'+r} \binom{k'+r}{k'} k' \mu^r (1-\mu)^{k'}}{\sum_{k'=0}^{\infty} p(k'+r) i_{k'+r} \binom{k'+r}{k'} \mu^r (1-\mu)^{k'}} \tag{S44}$$

where $i_k(f) \sim h^k$ is a function of $f$ representing the probability that a node of degree $k$ remains when a fraction $f$ of nodes has been removed. Here, $\mu$ is the probability that a neighbor of a node has been removed.

$$\begin{aligned} k_f(r) &= \frac{\sum_{k'=0}^{\infty} p(k'+r)\binom{k'+r}{k'} k'[(1-\mu)h]^{k'}}{\sum_{k'=0}^{\infty} p(k'+r)\binom{k'+r}{k'}[(1-\mu)h]^{k'}} \\ &= \frac{G_r^{(r+1)}\big(h(1-\mu)\big)}{G_r^{(r)}\big(h(1-\mu)\big)} h(1-\mu) \end{aligned} \tag{S45}$$

Combining equations ( S7) and (S12), we have:

$$\mu = 1 - \frac{G_1(h)}{h}\frac{1-f}{G_0(h)}. \tag{S46}$$

Since only functions of $f$ are replaced in this equation, the asymptotic behavior of $k_f$ depends solely on the network topology.

Therefore, we obtain:

$$k_f(r)\Big|_{r\to\infty} = W(h)H(r) \sim H(r), \tag{S47}$$

where

$$W(h) = \begin{cases} \frac{(1-f)G_1(h)}{G_0(h)-(1-f)G_1(h)} & \text{if } H(r)|_{r\to\infty} \to \infty, \\ (1-f)G_1(h)/G_0(h) & \text{if } H(r)|_{r\to\infty} = H_0, \end{cases} \tag{S48}$$

and $h$ is a function of $f$.

We have again:

$$k_f(r)\Big|_{r\to\infty} \sim H(r) = (r+1)\frac{p(r+1)}{p(r)}. \tag{S49}$$

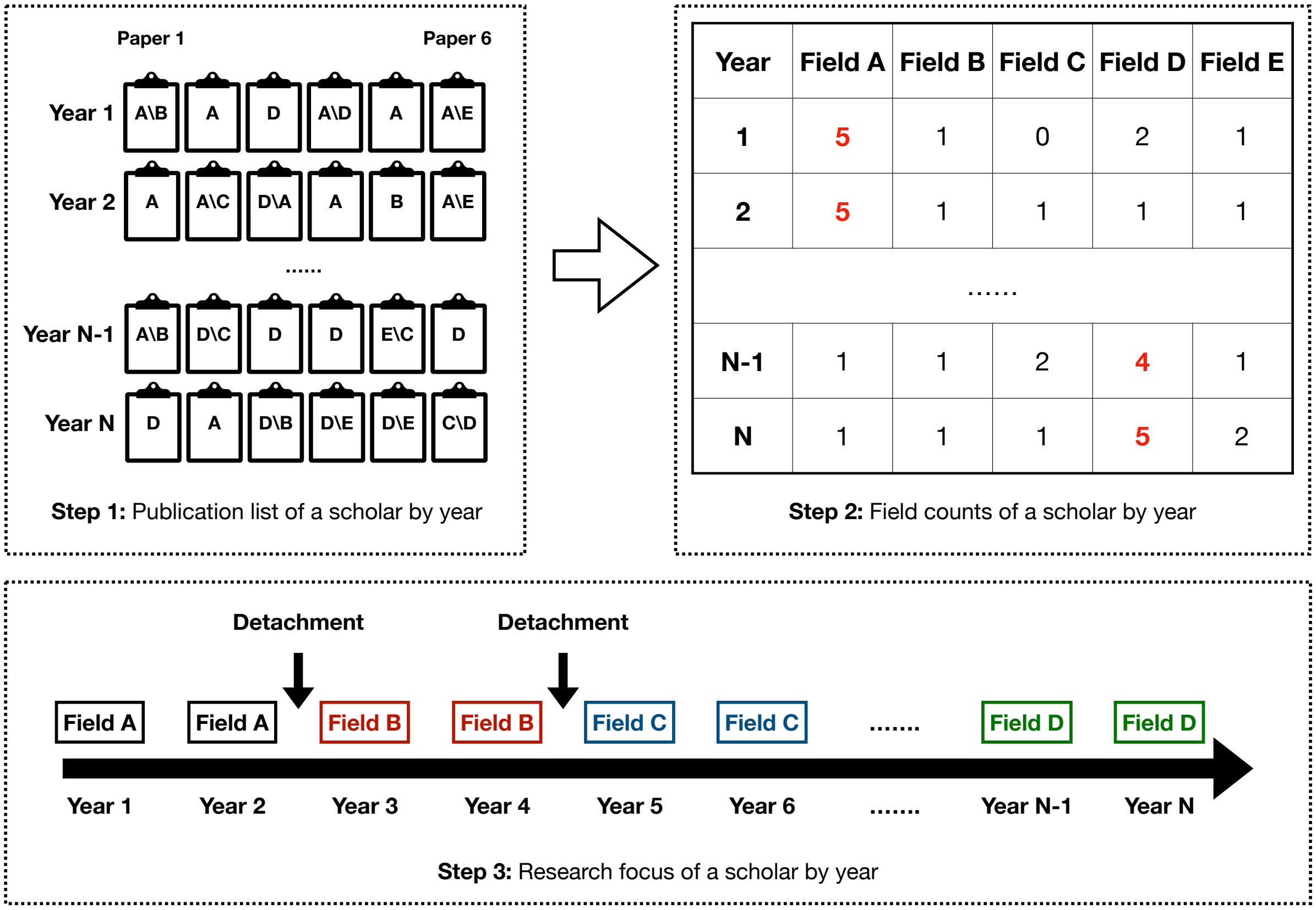

Figure S1: **Example of a scientist's field detachments.**

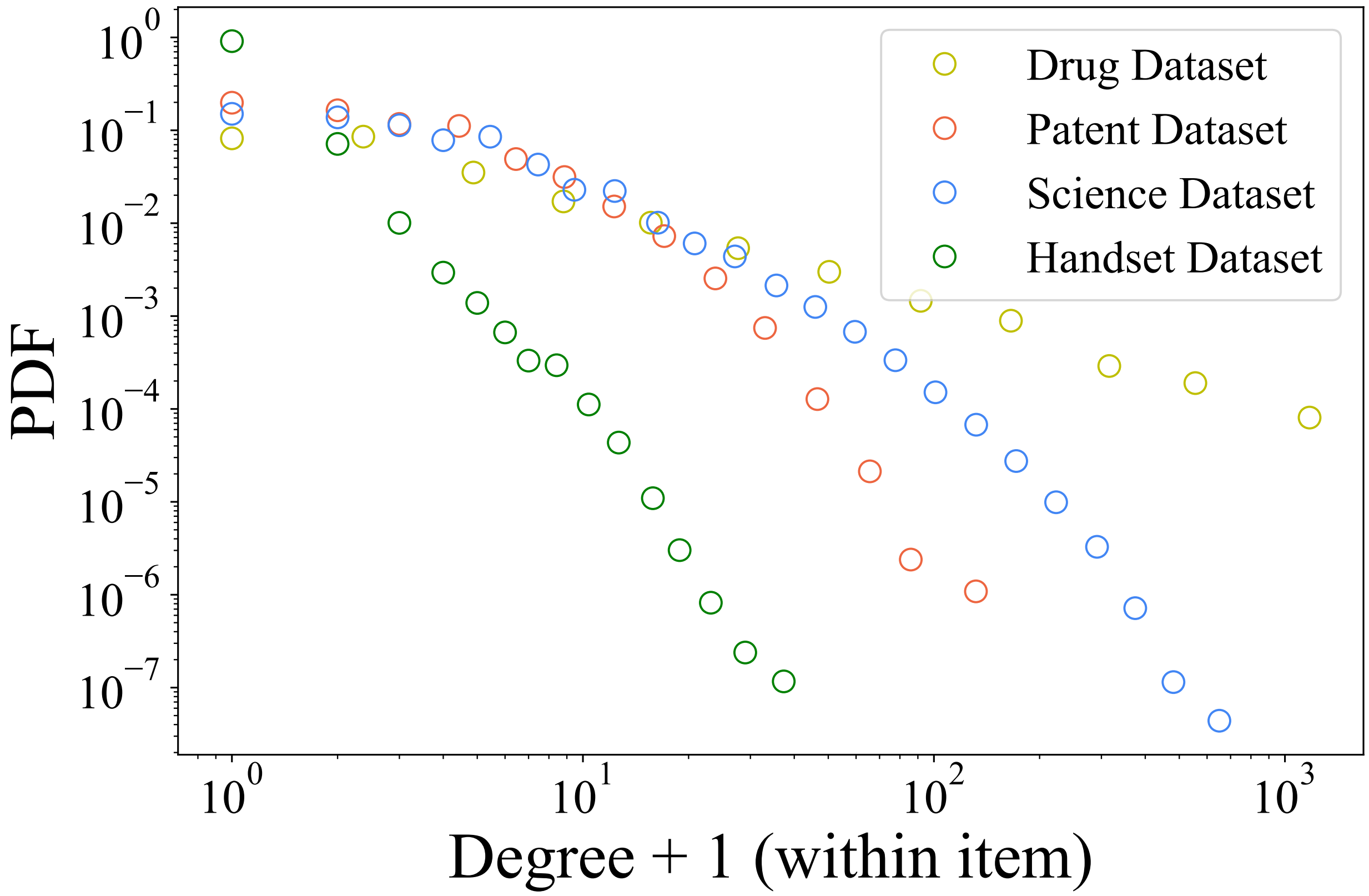

Figure S2: **Degree distributions of the empirical item-specific networks.** The figure plots the degree distributions of item-specific relational networks across the four domains: scientific fields, patenting domains, mobile handsets, and drug MOAs. In each domain, networks are constructed among actors associated with the same focal item using the corresponding underlying relation, including collaboration, communication, or market-focus similarity. The degree distributions are heavy-tailed across all four settings, indicating substantial heterogeneity in local connectivity and motivating the explicit treatment of network topology in the analysis of detachment dynamics.

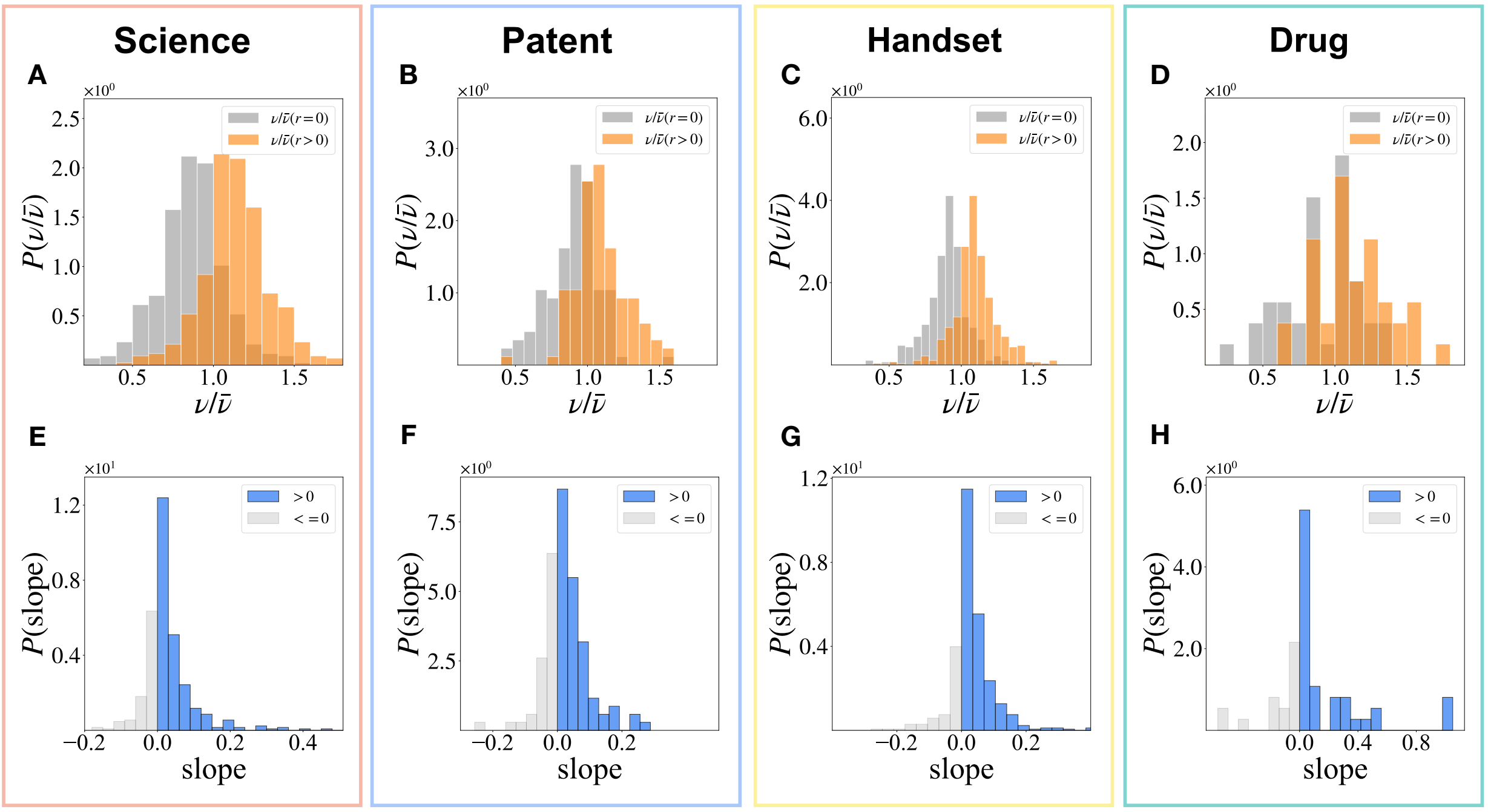

Figure S3: **Robustness check across different time intervals (reduced to 60% of maximum).**

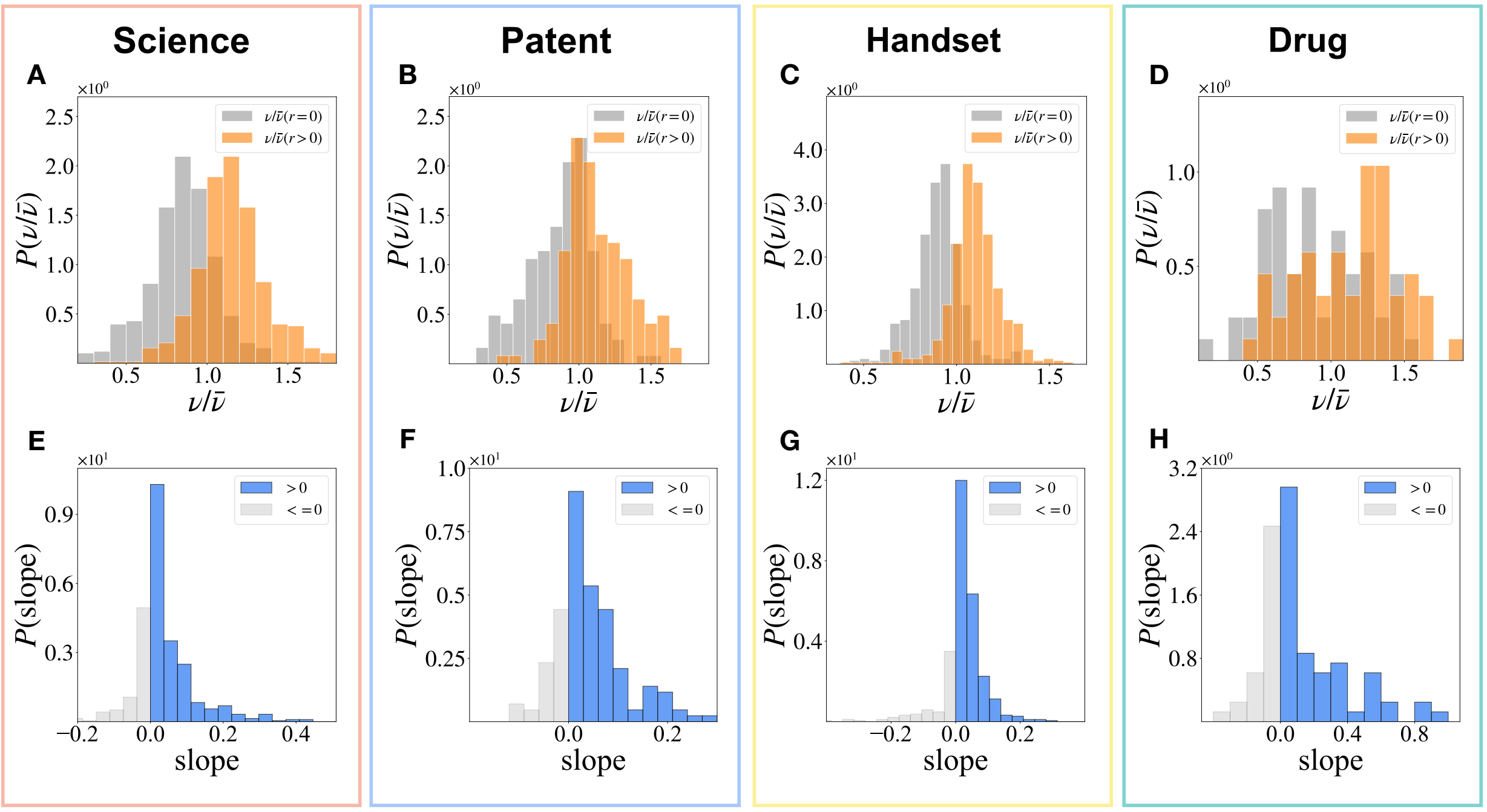

Figure S4: **Robustness check across different time intervals (reduced to 40% of maximum).**

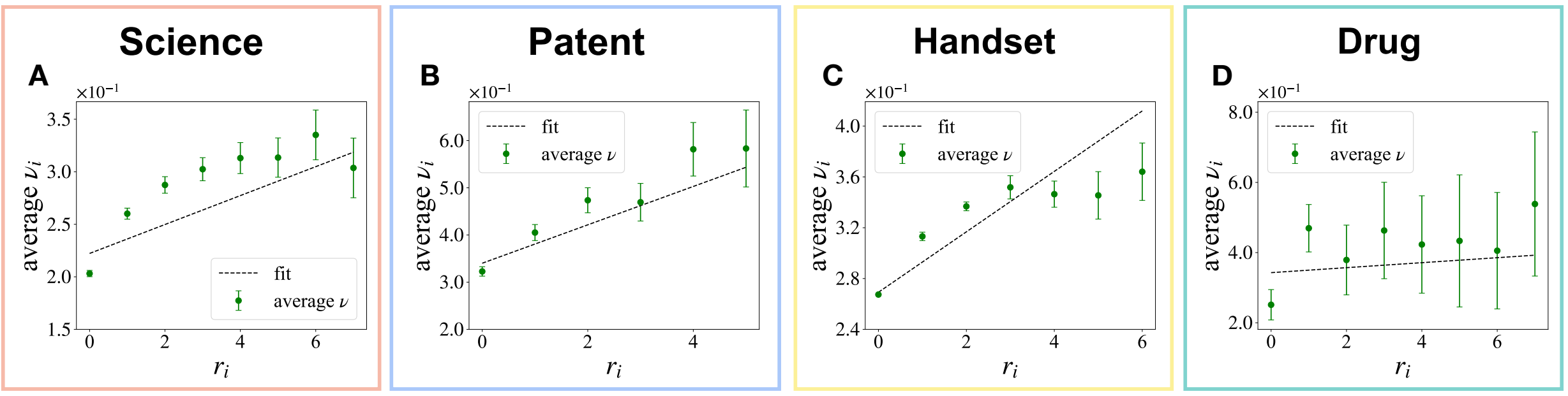

Figure S5: **Averaged detachment curves across the four empirical domains.** In addition to the representative item-level examples shown in Fig. 2E–H of the main text, we calculate averaged curves showing the relationship between detachment probability and the number of detached neighbors for each domain. The averaged curves provide a population-level visualization and show patterns consistent with the main empirical results in Fig. 2.

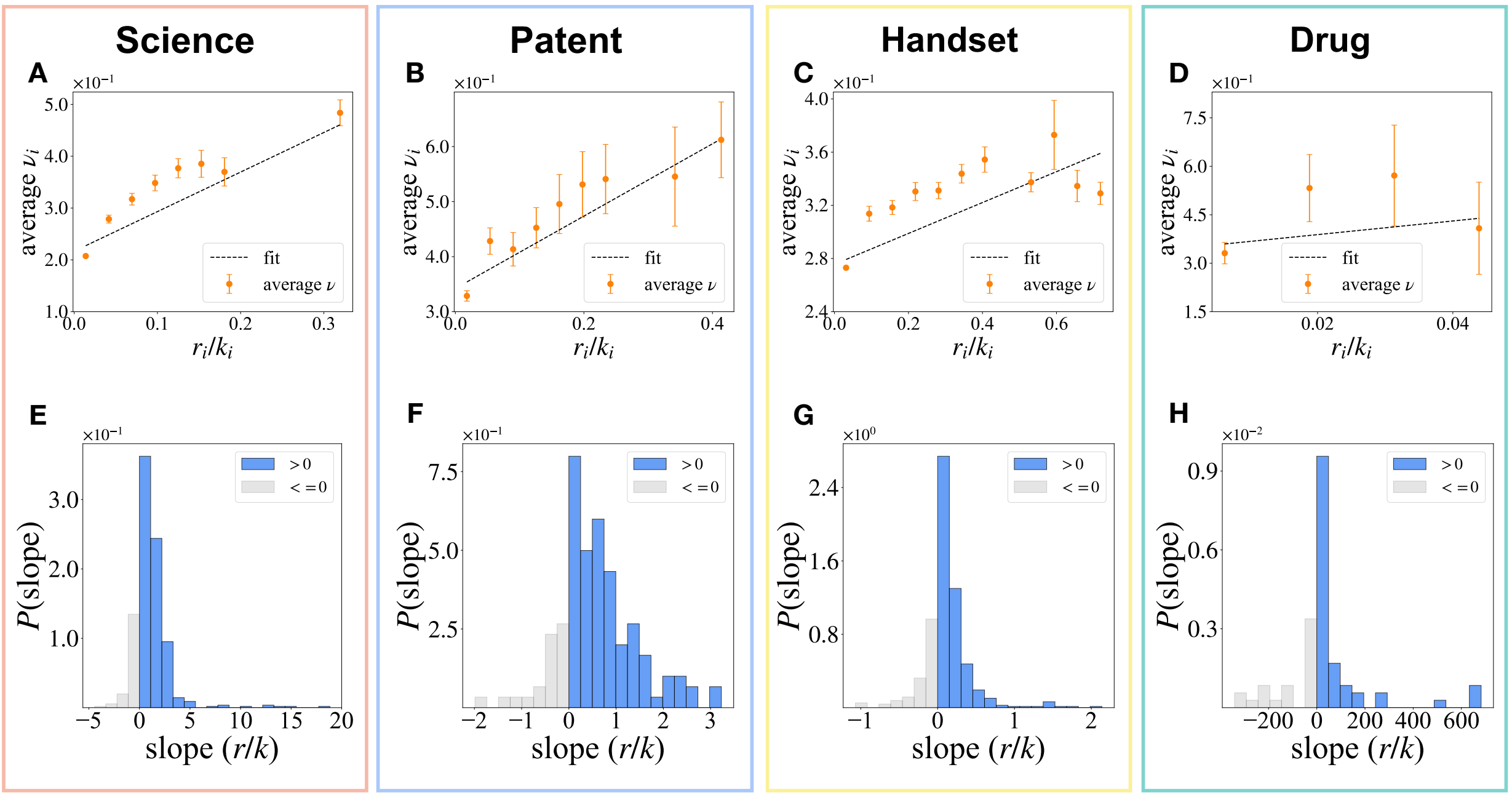

Figure S6: **Robustness of preferential detachment using normalized exposure.** We repeat the main empirical analysis using normalized exposure, $r_i/k_i$, defined as the fraction of node $i$'s neighbors who have already detached from the focal item. Across the four empirical domains, the detachment probability $\nu_i$ increases with normalized exposure, confirming that the observed preferential-detachment pattern is not solely driven by the raw number of detached neighbors.

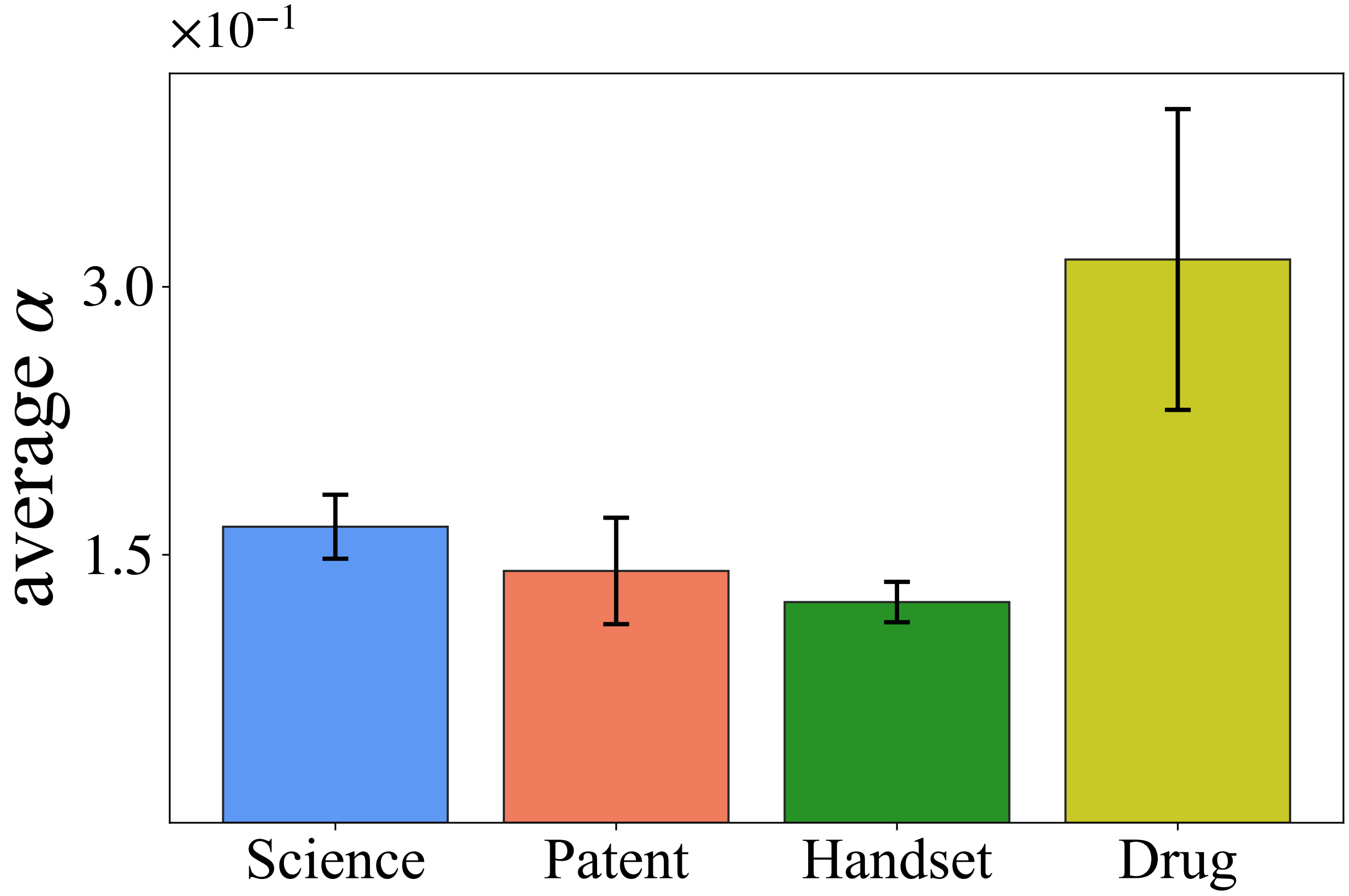

Figure S7: **Empirical estimates of preferential-detachment strength across domains.** We estimate an item-level empirical analogue of $\alpha$ from the fitted relationship between detachment probability and the number of detached neighbors. The figure reports the average estimated $\alpha$ values across the four empirical domains, with error bars indicating 95% confidence intervals.

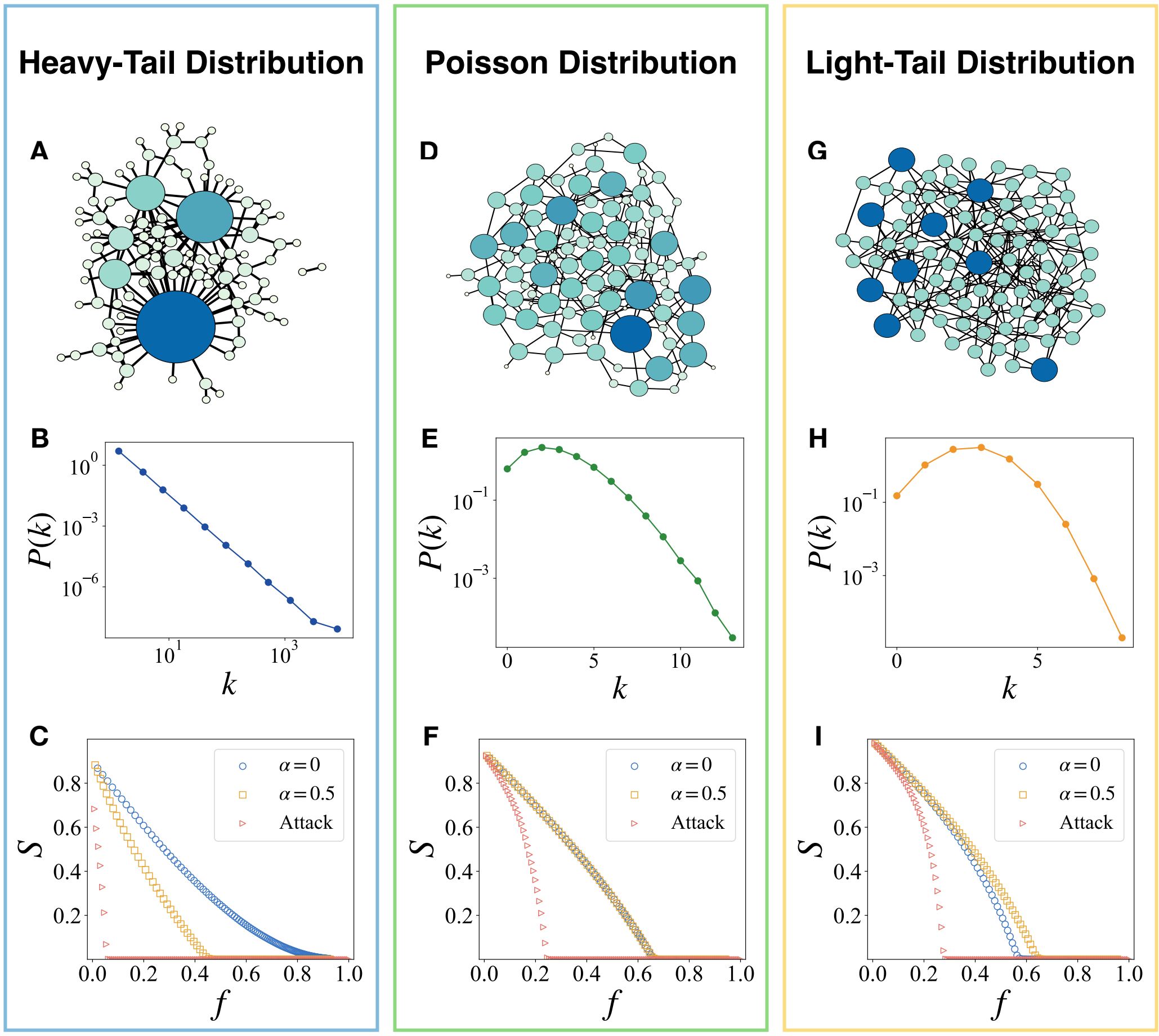

Figure S8: **Percolation process in different network topologies.**

# S5 Supplementary References

1. Sinha, A. *et al.* An overview of Microsoft Academic Service (MAS) and applications. In *Proceedings of the 24th International Conference on World Wide Web*, 243–246 (ACM, 2015).

2. Wang, K. *et al.* Microsoft Academic Graph: When experts are not enough. *Quantitative Science Studies* **1**, 396–413 (2020).

3. Barabási, A.-L. *Network Science* (Cambridge University Press, 2016).

4. Pastor-Satorras, R., Castellano, C., Van Mieghem, P. & Vespignani, A. Epidemic processes in complex networks. *Reviews of Modern Physics* **87**, 925 (2015).

5. Newman, M. *Networks* (Oxford University Press, 2018).

6. Molloy, M. & Reed, B. A critical point for random graphs with a given degree sequence. *Random Structures & Algorithms* **6**, 161–180 (1995).